\documentclass[%
 reprint,
superscriptaddress,
 amsmath,amssymb,
 aps,
]{revtex4-2}

\usepackage{graphicx}
\usepackage{dcolumn}
\usepackage{bm}

\begin{document}

\preprint{APS/123-QED}

\title{Hopper flows of deformable particles}

\author{Yuxuan Cheng}
 \affiliation{Department of Physics, Yale University, New Haven, Connecticut, 06520, USA.}
 \email{yuxuan.cheng@yale.edu}
\author{John D. Treado}
\affiliation{
 Department of Mechanical Engineering and Materials Science, Yale University, New Haven, Connecticut, 06520, USA.
}
\author{Ben Lonial}
\affiliation{
Department of Physics, Emory University, Atlanta, GA 30322, USA.
}
\author{Piotr Habdas}
\affiliation{
Department of Physics, Saint Joseph's University, Philadelphia, PA 19131, USA.
}
\author{Eric R. Weeks}
\affiliation{
Department of Physics, Emory University, Atlanta, GA 30322, USA.
}
\author{Mark D. Shattuck}
\affiliation{
Benjamin Levich Institute and Physics Department, The City College of New York, New York, New York 10031, USA.
}
\author{Corey S. O'Hern}
\affiliation{
Department of Mechanical Engineering and Materials Science, Yale University, New Haven, Connecticut, 06520, USA.
}
\affiliation{
Department of Physics, Yale University, New Haven, Connecticut, 06520, USA.
}
\affiliation{
Program in Computational Biology and Bioinformatics, Yale University, New Haven, Connecticut, 06520, USA
}

\date{\today}

\begin{abstract}
Numerous experimental and computational studies show that continuous hopper flows of granular materials obey the Beverloo equation that relates the volume flow rate $Q$ and the orifice width $w$: $Q \sim (w/\sigma_{\rm avg}-k)^{\beta}$, where $\sigma_{\rm avg}$ is the average particle diameter, $k\sigma_{\rm avg}$ is an offset where $Q\sim 0$, the power-law scaling exponent $\beta=d-1/2$, and $d$ is the spatial dimension.  Recent studies of hopper flows of deformable particles in different background fluids suggest that the particle stiffness and dissipation mechanism can also strongly affect the power-law scaling exponent $\beta$. We carry out computational studies of hopper flows of deformable particles with both kinetic friction and background fluid dissipation in two and three dimensions. We show that the exponent $\beta$ varies continuously with the ratio of the viscous drag to the kinetic friction coefficient, $\lambda=\zeta/\mu$. $\beta = d-1/2$ in the $\lambda \rightarrow 0$ limit and $d-3/2$ in the $\lambda \rightarrow \infty$ limit, with a midpoint $\lambda_c$ that depends on the hopper opening angle $\theta_w$.  We also characterize the spatial structure of the flows and associate changes in spatial structure of the hopper flows to changes in the exponent $\beta$. The offset $k$ increases with particle stiffness until $k \sim k_{\rm max}$ in the hard-particle limit, where $k_{\rm max} \sim 3.5$ is larger for $\lambda \rightarrow \infty$ compared to that for $\lambda \rightarrow 0$. Finally, we show that the simulations of hopper flows of deformable particles in the $\lambda \rightarrow \infty$ limit recapitulate the experimental results for quasi-2D hopper flows of oil droplets in water. 
\end{abstract}

\maketitle


\section{Introduction}

Silos and hoppers are used frequently in the agriculture \cite{Karimi2019}, pharmaceutical\cite{Faqih2007}, and consumer products industries\cite{Fitzpatrick2004} to
store fluids and granular materials. Materials confined within silos and hoppers are discharged using vertical or slanted walls that lead to an orifice at the bottom of the device. Microfluidic devices also incorporate flow constrictions to control the pressure and flow rate of complex fluids, such as emulsion droplets\cite{Bick2021}.  Despite the fact that hopper and silo flows are ubiquitous in industry, we do not yet have a fundamental understanding of the outflow properties from hoppers and silos. For example, it is difficult to predict the outflow rate of particulate materials from hoppers and silos as a function of the device geometry, orifice size, and particle properties.    

For inviscid fluid flows from hoppers, the volume flow rate $Q$ is proportional to the orifice area ($w^2$ in three dimensions, where $w$ is the diameter of the circular orifice) times the characteristic fluid velocity $v_c$ at the orifice, $Q = w^2 v_c$\cite{Alessio2021}. For pressure-driven flows, $v_c \sim \sqrt{\Delta P/\rho}$, where $\Delta P$ is the pressure difference and $\rho$ is the mass density of the fluid. For viscous fluid flows, the volume flow rate $Q = C_d w^2 v_c$ includes a discharge coefficient $C_d$ that depends on the hopper geometry and viscosity of the fluid\cite{Essien2019}. 

Unlike ordinary fluids, granular materials consist of macro-sized grains that interact via dissipative forces, which can give rise to intermittency and clogging during hopper flows in the limit of small orifice sizes. Beverloo and co-workers\cite{BEVERLOO1961260} carried out seminal experimental studies of hopper flows of a wide range of granular materials in air and proposed an empirical form for the flow rate that allows flow arrest to occur at nonzero orifice width:
\begin{equation}
\label{eq:1}
Q(w) = C(w/\sigma_{\rm avg}-k)^{\beta},
\end{equation}
where $C$ is a constant with units of flow rate, $\sigma_{\rm avg}$ is the average diameter of the particles, $Q(k \sigma_{\rm avg})=0$, and $k$ depends on the particle properties, such as the stiffness, shape, and friction coefficient. Another key difference between hopper flows of ordinary fluids and granular materials is that the power-law scaling exponent $\beta = d-1/2$, where $d$ is the spatial dimension, is not an integer for hopper flows of granular materials. This relation has been verified in 2D and 3D, for spherical \cite{Pascot2020,Hirshfeld1997} and non-spherical \cite{Tang2016} particles, and for frictionless\cite{Ashour2017} and frictional\cite{Sheldon2010} particles. 

Numerous researchers have provided heuristic arguments for the $\beta=d-1/2$ scaling exponent for hopper flows of granular materials. For example, Brown and Richards proposed a model for the regime $w\gg k\sigma_{\rm avg}$ where transient arches form and break in a region above the orifice, creating a free-fall region below with a height proportional to the orifice width $w$ \cite{Brown1961}. Because of shielding by the transient arches, the grains move at low velocities until they enter the free-fall region. Thus, the discharge velocity $v_{c} \sim \sqrt{gw}$ when grains reach the orifice, and $Q \sim w^2 v_c \sim w^{5/2}$ in 3D or $Q\sim w v_c \sim w^{3/2}$ in 2D. Cutoffs for the finite size of the particles can be added to these expressions to recover Eq.~\ref{eq:1}.

The original studies of Beverloo {\it et al.} involved hopper flows of hard grains in air\cite{BEVERLOO1961260}. Recent studies of hopper flows of spherical glass beads submerged in water have found that the scaling exponent $\beta \sim 1$ does not obey $\beta = d -1/2$ from the original Beverloo equation\cite{Wilson2014,Fan2022}. In addition, studies of qausi-2D hopper flows of air bubbles immersed in water have found $\beta \sim 0.5$\cite{Bertho2006}, again deviating from the exponent in the original Beverloo equation. Thus, from these previous results, it is not clear whether the dissipation mechanism (i.e. particle-particle or background fluid dissipation), particle stiffness or other particle properties control the power-law scaling exponent in Eq.~\ref{eq:1}. 

In this article, we carry out computer simulations of hopper flows of deformable particles in two (2D) and three dimensions (3D), including both interparticle kinetic friction and viscous dissipation with the background fluid. We employ two computational models of particle deformation: 1) the ``soft particle" model that describes particle deformation as overlaps between pairs of particles and therefore does not conserve particle volume in 3D (area in 2D) and 2) the deformable particle model that includes a shape-energy function for changes in particle volume (area in 2D), surface area (perimeter in 2D), and surface bending, as well as an interaction energy that prevents particle overlaps. Studying these two models allows us to assess the importance of volume conservation in determining the flow properties and provides the ability to tune the particle stiffness, static and kinetic friction coefficients, and background viscous drag and quantify their effects on the flow rate.  

We find several important results. First, the power-law scaling exponent $\beta$ relating the volume flow rate $Q$ and orifice width $w$ is controlled by the dissipation mechanism, {\it i.e.}~the ratio of the viscous damping coefficient to the kinetic friction coefficient, $\lambda=\zeta/\mu$. We find that the exponent varies continuously between $\beta = d-1/2$ in the $\lambda \rightarrow 0$ limit and $d-3/2$ in the $\lambda \rightarrow \infty$ limit, with a midpoint $\lambda_c$ that depends on the hopper opening angle $\theta_w$. In contrast, the exponent $\beta$ is only weakly dependent on the particle deformability and surface roughness. Second, we show that the spatio-temporal dynamics for flows with the two exponents, $\beta = d-1/2$ and $d-3/2$, are different. In particular, the velocity profile varies more strongly with the orifice size for flows with $\beta = d-1/2$ in the $\lambda \rightarrow \infty$ limit. Third, the offset $k\sigma_{\rm avg}$ at which $Q \rightarrow 0$ decreases with particle deformability, and increases with the static friction coefficient.  Finally, we show that the simulations of hopper flows using the soft and deformable particle models in the $\lambda \rightarrow \infty$ limit are able to recapitulate the experimental results for quasi-2D gravity-driven hopper flows of oil droplets in water. 

The remainder of the article is organized as follows. In Section~\ref{sim_methods}, we describe the simulation methods including the soft particle and deformable particle models, the equations of motion, and simulation protocol that we employ to generate continuous flows. In Section~\ref{expmethods}, we describe the experimental system, including the hopper geometry and method to generate emulsion droplets and flows. In Section \ref{results}, we show results for the volume flow rate (area flow rate in 2D) $Q$ versus the orifice width $w$ for the soft particle model and the deformable particle model as a function of $\zeta/\mu$ and particle deformability in both 2D and 3D. We characterize the spatial structure of the flows by measuring the velocity as a function of distance from the orifice and we associate changes in the spatial structure of the flows to changes in the power-law scaling exponent $\beta$. In Section 4, we discuss the implications of our results, and propose future research directions, such as developing an improved deformable particle model that includes surface tension, which would allow more quantitative comparisons between the simulations and experiments on hopper flows of oil droplets in water. We also include three Appendices. In Appendix A, we describe the details of the frictionless, deformable particle model. In Appendix B, we show more detailed comparisons of the flow rate for the soft particle and deformable particle models in the compressible and incompressible particle limits.  In Appendix C, we show that the system size effects on the flow rate are small in the simulations.

\begin{figure*}[!ht]
\includegraphics[width=0.98\textwidth]{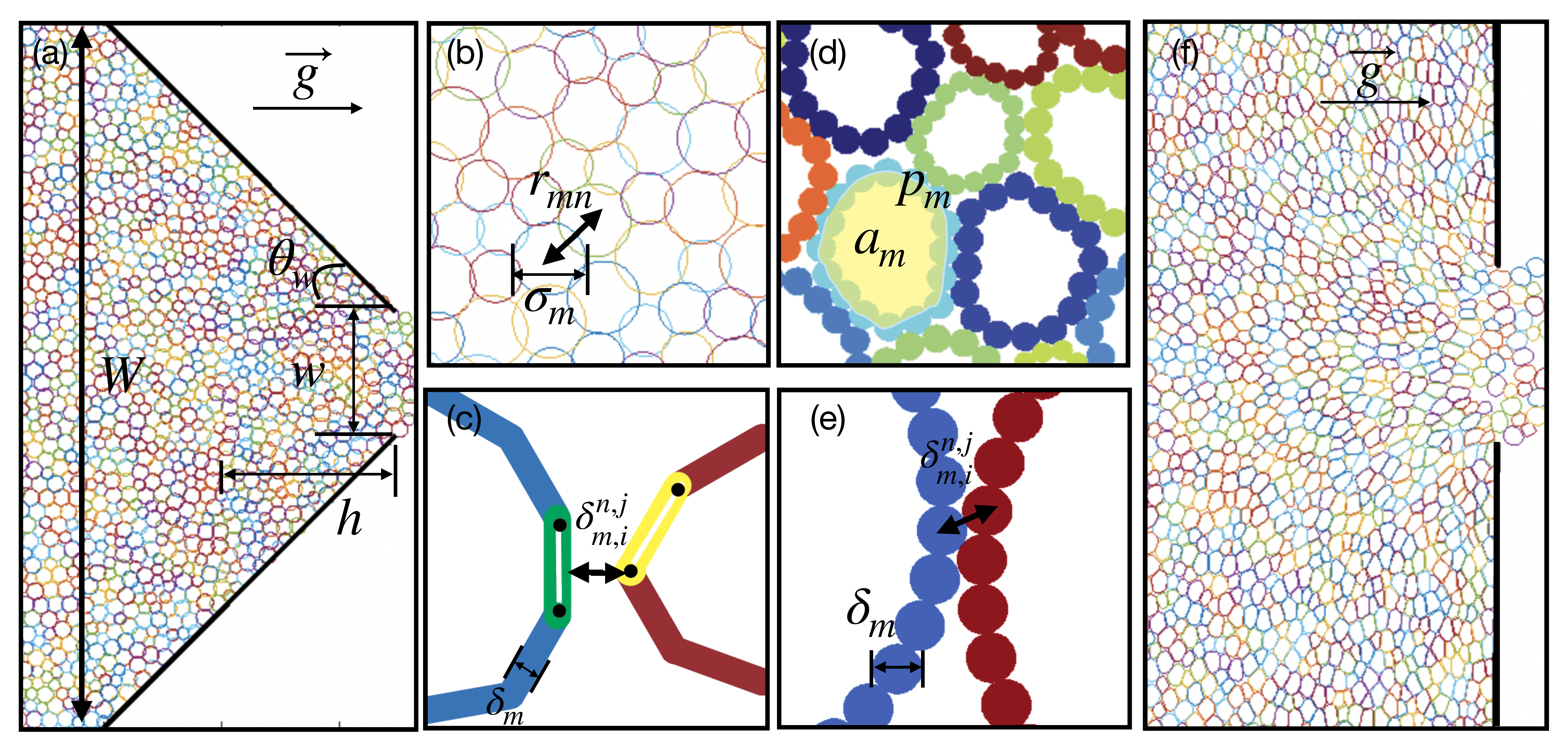}
\centering
\caption{Snapshot from simulations of hopper flows of bidisperse mixtures in a gravitational field using the (a) soft particle (SP) and (f) deformable particle (DP) models in 2D.  The hopper geometry can be slanted with variable tilt angle $\theta_w$, e.g. $\theta_w= 45^\circ$ in (a) and $90^\circ$ in (f). ${\vec g}$ indicates the direction of the gravitational acceleration, $W$ is the separation between the straight walls far from the orifice, $h$ indicates the distance from the hopper orifice, and $w$ is the width of the orifice. (b) Close-up of hopper flow using the SP model with $N/2$ large particles and $N/2$ small particles with diameter ratio $1.4$, highlighting overlapping particles $m$ and $n$ with separation $r_{mn} < \sigma_{mn}$, where $\sigma_{mn} = (\sigma_m+\sigma_n)/2$. (c) Illustration of the method to calculate the closest separation between frictionless, deformable particles $m$ and $n$. $\delta_m$ is the width of the edges of particle $m$ and $\delta_{m,i}^{n,j}$ is the shortest distance between edges $i$ and $j$ on particles $m$ and $n$, respectively. (See Appendix A.) (d) Close-up of hopper flow using the DP model with surface roughness with $N/2$ large particles, $N/2$ small particles, and area ratio $1.96$. $a_m$ is the area and $p_m$ is the perimeter of deformable particle $m$. 
Both small and large particles have $N_v =16$ vertices.
(e) Illustration of the interactions between deformable particles $m$ and $n$ with surface roughness. $\delta_m$ is the diameter of each circular vertex on particle $m$ and $\delta_{m,i}^{n,j}$ is the distance between vertices $i$ and $j$ on particles $m$ and $n$, respectively. } 
    \label{fig:1}
\end{figure*}

\section{Simulation Methods}
\label{sim_methods}

In this section, we describe the methods for simulating gravity-driven hopper flows of bidisperse particles in 2D and 3D.  We first illustrate the hopper geometry. We then describe the two methods for modeling the particle shape and interactions: 1) the soft particle model, which treats each spherical particle as a single degree of freedom located at its center of mass and mimics particle contact interactions by allowing overlaps between pairs of particles and 2) the deformable particle model that uses a shape-energy function to penalize changes in particle volume (area in 2D), surface area (perimeter in 2D), and surface bending. The deformable particle model can be implemented such that the particles are nearly frictionless or the model can include surface roughness. For each model, we describe the forces that result from the shape-energy function, particle-particle interactions, and dissipative forces arising from interparticle kinetic friction and drag from the background fluid, and then we write down the resulting equations of motion for each particle.  Finally, we discuss the initialization of the particle positions and velocities and the method used to generate continuous flows.      

\subsection{Hopper Geometry}

In 2D, the hopper is constructed from two infinitely long straight (top and bottom) walls separated by a distance $W \sim 60\sigma_s$ (where $\sigma_s$ is the diameter of the small particle), which connect to the right wall at an angle $\theta_w$ as shown in Fig.\ref{fig:1} (a). The gravitational field points from left to right. The orifice is centered and has width $w<12\sigma_s$, so that $W/w > 5$, which ensures that the top and bottom walls are sufficiently separated such that they do not influence the flow. In 3D, the hopper is an infinitely long cylinder with diameter $W\sim 30\sigma_s$, and the long axis of the cylinder is oriented in the direction of gravity. The hopper in 3D has a flat base ($\theta_w=90^\circ$) containing a circular orifice with diameter $w$ that is centered on the long axis of the cylinder.

In 2D, we focus on systems containing $N=1600$ particles, but we also considered systems over a range from $N=800$ to $3200$ to assess system size effects. In 3D, we focus on systems with $N=6400$ particles. To mimic continuous flows, particles that exit the hopper orifice are replaced on the left side of the hopper near the leftmost flowing particles and given the same speed as neighboring particles. The distance between the hopper orifice and the leftmost flowing particle is $L \sim 20$-$30\sigma_s$.

\subsection{Soft Particle Model}

For gravity-driven hopper flows, there are typically four contributions to the total potential energy: 1) the shape-energy function $U_m^s$, 2) the gravitational potential energy $U_m^g$, 3) the particle-particle interaction energy $U_{\rm int}$, and 4) the particle-wall interaction energy $U_m^w$. For the SP model, $U_m^s=0$. Purely repulsive interparticle forces are generated by allowing overlaps between pairs of spherical particles\cite{Durian1995,Durian1997,Hong2017,Tao2021}, as shown in Fig.\ref{fig:1} (b). The pairwise interaction energy of the SP model is given by
\begin{equation}
\label{sp}
U_{\rm int} = \sum_{m=1}^{N}\sum_{n>m}^{N} \frac{\epsilon_{sp}}{2} (1-r_{mn}/\sigma_{mn})^2\Theta(1-r_{mn}/\sigma_{mn}).
\end{equation}
In Eq.~\ref{sp}, $\sigma_{mn} = (\sigma_m + \sigma_n)/2$ is the average diameter of particles $m$ and $n$, $r_{mn}$ is the separation between particles $m$ to $n$, and $\epsilon_{sp}$ is the characteristic energy scale of the repulsive interaction. The Heaviside step function $\Theta(\cdot)$ ensures that the pair forces are non-zero only between overlapping particles. 

We consider a similar repulsive interaction between the hopper walls and each particle $m$ that is in contact with the walls:
\begin{equation}
U_{m}^{w} =\frac{\epsilon_{w}}{2} (1-2d_w/\sigma_{m})^2\Theta(1-2d_w/\sigma_{m}),
\end{equation}
where $d_w$ is the distance between the center of particle $m$ and the hopper wall and $\epsilon_w$ is the characteristic energy scale of the particle-wall interaction. Thus, the total potential energy of the system is given by
\begin{equation}
\label{totalU}
U = \sum_{m=1}^{N} (U_{m}^{s} + U_{m}^{g} +U_{m}^{w}) + U_{\rm int},
\end{equation}
where $U_{m}^{g} = -M_m g h$, $h$ is the height of the center of mass of particle $m$, $g$ is the gravitational acceleration, $M_m = \rho V_{m,0}$ is the mass of particle $m$ with mass density $\rho$ and volume $V_{m,0} = \pi \sigma_m^3/6$.
(In 2D,  $M_m = \rho a_{m,0}$ is the mass of particle $m$ with areal mass density $\rho$ and area $a_{m,0} = \pi \sigma_m^2/4$.)

We include two types of dissipative forces on the particles. First, we consider viscous drag forces on particles moving in a background viscous fluid:
\begin{equation}
\vec{F}_{m}^{\zeta} = - \zeta \vec{v}_m,
\end{equation}
where $\zeta$ is the drag coefficient and $\vec{v}_m$ is the velocity of particle $m$. The second dissipative force arises from kinetic friction between contacting particles. The kinetic friction force is proportional to the relative velocity between contacting particles\cite{Silbert2002}:
\begin{equation}
\vec{F}_{m}^{\mu} = - \mu  \sum_{n\neq m}^{N}  (\vec{v}_m  - \vec{v}_n )\Theta(1-r_{mn}/\sigma_{mn}),
\end{equation}
where $\mu$ is the kinetic friction coefficient. The dimensionless parameter $\lambda =\zeta/\mu$ determines whether the energy dissipation arises mainly from viscous drag ($\lambda \gg 1$) or from kinetic friction ($\lambda \ll 1$). We measure the kinetic friction and drag coefficients in units of $\mu_0 = \zeta_0 = \rho \sigma_{\rm avg}^{d-1} g t_0$, where $t_0 = \sqrt{\sigma_{\rm avg}/g}$. For the SP model, the equation of motion for each particle $m$ is
\begin{equation}
\label{eom}
M_m \frac{\partial^2 \vec{r}_m}{\partial t^2} = -\vec{\nabla}_{r_m}U + \vec{F}_{m}^{\zeta}  + \vec{F}_{m}^{\mu}.
\end{equation}
We integrate Eq.~\ref{eom} using a modified velocity Verlet integration scheme with time step $\Delta t=10^{-3} t_0$.  The flow rate $Q$ is measured in units of $Q_0 = \sigma_{\rm avg}^d/t_0$.

For the SP model, we focus on bidisperse systems in 2D and 3D composed of half large particles and half small particles with diameter ratio $\alpha = \sigma_l/\sigma_s = 1.4$ to avoid crystallization\cite{Zhang2013}. The average diameter of particles in the bidisperse system is $\sigma_{avg} = (\sigma_l+\sigma_s)/2 = 1.2\sigma_s$. Two important dimensionless energy scales are the ratios of the characteristic particle-particle and particle-wall repulsive energy scales to the gravitational potential energy, i.e. $E_{sp} = \epsilon_{sp}/(g\rho \sigma_{\rm avg}^{d+1})$ and $E_w = \epsilon_{w}/(g\rho \sigma_{\rm avg}^{d+1})$, where $d=2$, $3$ in two and three dimensions, respectively. We set $E_w = 10^4$ to minimize overlaps between the particles and hopper walls and will vary $E_{sp}$ to determine the effect of particle softness on the flow rate $Q(w)$.

\subsection{Deformable Particle Model}

To explicitly model changes in particle shape, we recently developed the deformable particle (DP) model in both 2D \cite{Boromand2018,Boromand2019} and 3D\cite{Wang2021}. In 2D, the particles are modeled as deformable polygons composed of $N_v$ vertices. We can achieve deformable particles with nearly smooth surfaces by modeling the vertices as circulo-lines as shown in Fig.\ref{fig:1} (c) or achieve deformable particles with nonzero surface roughness by modeling the vertices as small disks as shown in Fig.\ref{fig:1} (d) and (e). We consider the following shape-energy function for particle $m$:

\begin{equation}\label{shape_energy}
\begin{split}
U_{m}^{s} =& \frac{k_a}{2} (a_m-a_{m,0})^2 + \frac{k_l N_v}{2}\sum_{i=1}^{N_v} (l_{m,i}-l_{m,0})^2 \\ &+
 \frac{k_b}{2N_v}\sum_{i=1}^{N_v}\left(\frac{\hat{l}_{m,i}-\hat{l}_{m,i+1}}{l_{m,0}}\right)^2, 
 \end{split}
\end{equation}

which includes three terms. The first term imposes a harmonic energy penalty for changes in particle area $a_m$ from the preferred value $a_{m,0}$ and $k_a$ controls the fluctuations in particle area. The second term imposes a harmonic energy penalty for deviations in the separations $l_{m,i}$ between adjacent vertices $i$ and $i+1$ from the equilibrium length $l_{m,0}$ and $k_l$ controls fluctuations in the separations between adjacent vertices. The third term is the bending energy that favors particle shapes with $\hat{l}_{m,i}$ and $\hat{l}_{m,i+1}$ in the same direction. $k_b$ is the bending rigidity that controls fluctuations in the angle between $\hat{l}_{m,i}$ and $\hat{l}_{m,i+1}$.  The factor of $N_v$ in the numerator of the second term and in the denominator of the third term of Eq.~\ref{shape_energy} ensure that $U_m^s$ does not depend on $N_v$. 

We focus on hopper flows of $N=1600$ bidisperse deformable particles in 2D with half large particles and half small particles.  We define effective diameters $\sigma_l = \sqrt{4a_{0,l}/\pi}$ and $\sigma_s = \sqrt{4a_{0,s}/\pi}$ for the large and small particles, respectively, and set the diameter ratio $\sigma_l / \sigma_s = 1.4$. We choose $N_v = 16$, which gives an effective friction coefficient $\mu_{\rm eff} \sim 0.6$ for the DP model with surface roughness\cite{Papanikolaou2013}.  For the nearly smooth DP model, we find that $N_v \ge 16$ does not affect the properties of the hopper flows. From $a_{0,s}$ and $l_{0,s}$, we can define the dimensionless shape parameter in 2D, ${\cal A}_0 = (N_v l_{0,s})^2/4\pi a_{0,s}$. We study systems composed of nearly circular particles with ${\cal A}_0 =(N_v/\pi)\tan (\pi/N_v) \sim 1.013$, which is the value for a regular polygon with $N_v=16$ sides. 

For the DP model with surface roughness, each vertex in particle $m$ is represented by a disk with diameter $\delta_m = l_{m,0}$ and the total interaction energy $U_{\rm int}$ is calculated by summing up all the repulsive interactions between overlapping circular vertices on different particles:
\begin{equation}\label{eq:9}
U_{\rm int} = \sum_{m=1}^{N}\sum_{n>m}^{N}\sum_{i=1}^{N_v}\sum_{j=1}^{N_v} \frac{\epsilon_{c}}{2} (1-\delta_{m,i}^{n,j}/\delta_{mn})^2\Theta(1-\delta_{m,i}^{n,j}/\delta_{mn}),
\end{equation}
where $\delta_{mn} = (\delta_m + \delta_n)/2$ is the average vertex diameter on particles $m$ and $n$, $\epsilon_c$ gives the characteristic energy scale of the repulsive interactions between vertices, and $\delta_{m,i}^{n,j}$ is the separation between vertex $i$ on particle $m$ and vertex $j$ on particle $n$. For the nearly smooth DP model, we represent edges of the polygon as circulo-lines with width $\delta_m = 0.1 l_{m,0}$ and length $l_{m,i}$. The interparticle repulsive interactions still follow Eq.\ref{eq:9}, but $\delta_{m,i}^{n,j}$ represents the distance between edges $i$ and $j$ on particles $m$ and $n$, respectively. See Appendix A for more details on implementing the nearly frictionless DP model in 2D.

The wall interaction between vertex $i$ on particle $m$ and the hopper wall is
\begin{equation}
\label{dp_wall}
U_{m,i}^{w} =\frac{\epsilon_{w}}{2} (1-2d_w/\delta_{m})^2\Theta(1-2d_w/\delta_{m}),
\end{equation}
where $d_w$ is the minimum distance between vertex $i$ on particle $m$ and the hopper wall. The total potential energy $U$ is again the sum of the shape-energy function $U_{m}^{s}$, the gravitational potential energy $U_{m}^{g}$, and the particle-wall interactions $U_{m}^{w}$ over all particles plus the potential energy from particle-particle interactions $U_{\rm int}$, as given in Eq.~\ref{totalU}.  

As for the SP model, we consider two types of dissipative forces acting on the deformable particles. Since we will write equations of motion for each vertex, we consider dissipative forces acting on the individual vertices. First, the viscous drag force on vertex $i$ on particle $m$ is:
\begin{equation}
\vec{F}_{m,i}^{\zeta} = - \frac{\zeta}{N_v} {\vec v}_{m,i},
\end{equation}
where ${\vec v}_{m,i}$ is the velocity of vertex $i$ on particle $m$. The kinetic friction force on vertex $i$ on particle $m$ arising from an overlap with vertex $j$ on particle $n$ is 
\begin{equation}
\vec{F}_{m,i}^{\mu} = - \mu  \sum_{n\neq m}^{N} \sum_{j=1}^{N_v} ({\vec v}_{m,i}  - {\vec v}_{n,j}) \Theta(1-\delta_{m,i}^{n,j}/\delta_{mn}).
\end{equation}
Thus, for the DP model, the equation of motion for vertex $i$ on particle $m$ is
\begin{equation}
M_{m,i} \frac{\partial^2 \vec{r}_{m,i}}{\partial t^2} = -\vec{\nabla}_{r_{m,i}}U + \vec{F}_{m,i}^{\zeta}  + \vec{F}_{m,i}^{\mu},
\end{equation}
where $M_{m,i} = M_m/N_v$ is the mass of vertex $i$ on particle $m$. 
From Eqs.~\ref{shape_energy},~\ref{eq:9}, and~\ref{dp_wall}, we can obtain five dimensionless energy scales for the DP model in a gravitational field in 2D: $K_a = k_a\sigma_{\rm avg}^2/(g\rho)$, $K_l = k_l/(g\rho)$, $K_{b} = k_{b}/(g\rho \sigma_{\rm avg}^4)$, $E_w = \epsilon_{w}/(g\rho \sigma_{\rm avg}^2)$ and $E_c = \epsilon_{c}/(g\rho \sigma_{\rm avg}^2)$, where $\sigma_{\rm avg}=(\sigma_s+\sigma_l)/2$. We choose $K_a>10^4$ so that the fluctuations in the particle areas are negligible. We also set $K_c = K_w = 10^4$ to minimize vertex-vertex and vertex-wall overlaps. We will vary $K_l$ and $K_b$ to determine their effects on the flow rate. The time $t$ and flow rate $Q$ are measured in units of $t_0 = \sqrt{\sigma_{\rm avg}/g}$ and $Q_0 = \sigma_{\rm avg}^d/t_0$. The equations of motion are integrated using a modified velocity Verlet algorithm with a time step of $10^{-3} t_0$. 

\subsection{Simulation Initialization}

For the DP model, we initialize the particles as regular polygons, and set the edge lengths to be equal to their equilibrium values $l_{m,0} = \sqrt{4a_{m,0} N_v \tan(\pi/N_v)}/N_v$. For both the SP and DP models, we randomly place the particles within the hopper with zero velocity. Initially, gravity is turned off, and energy minimization (using FIRE\cite{Bitzek2006}) is carried out to ensure no overlaps between the particles and the particles and the walls. After the removal of overlaps, gravity is turned on and the particles begin to fall toward the orifice. To achieve continuous flow, particles that exit the hopper orifice are placed back into the left side of the hopper in contact with one of the bulk particles with the same velocity as the particle it is touching.  A particle is considered outside of the hopper (and does not contribute to the flow rate) when it first exits the orifice. However, particles are put back into the hopper only after they fall at least two particle diameters past the orifice. 

\begin{figure}[!tp]
\centering
\includegraphics[width=0.8\columnwidth]{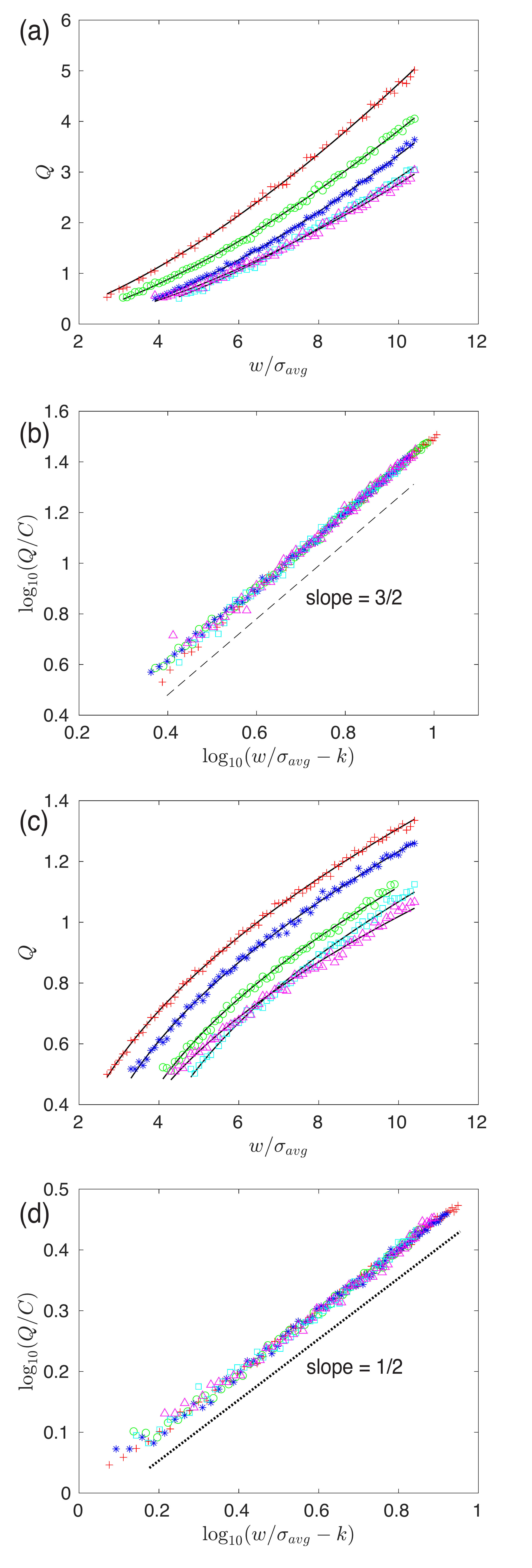}
\centering
\caption{Area flow rate $Q$ versus orifice width $w/\sigma_{\rm avg}$ for hopper flows in 2D using the SP and DP models with (a) kinetic friction only, $\mu/\mu_0 = 10\sqrt{10}$, and (c) viscous drag only, $\zeta/\zeta_0 = 1/\sqrt{10}$. We consider the SP model with $E_{sp}=10^2$ (asterisks) and $10^4$ (squares), the frictionless DP model with $K_l=10$ and $K_b=10^{-1}$ (crosses) and $K_l=10$ and $K_b=10$ (circles), and the DP model with surface roughness with $K_l=10$ and $K_b =10^{-1}$ (triangles). The solid curves in (a) and (c) are fits to the power-law scaling relation in Eq.~\ref{eq:1}. In (b) and (d), we show $\log_{10}(Q/C)$ versus $\log_{10}(w/\sigma_{\rm avg}-k)$ for the data in (a) and (c), and the dotted and dashed lines have slopes of $1/2$ and $3/2$, respectively.}
\label{fig:2}
\end{figure}

\begin{figure}[!ht]
\includegraphics[width=0.95\columnwidth]{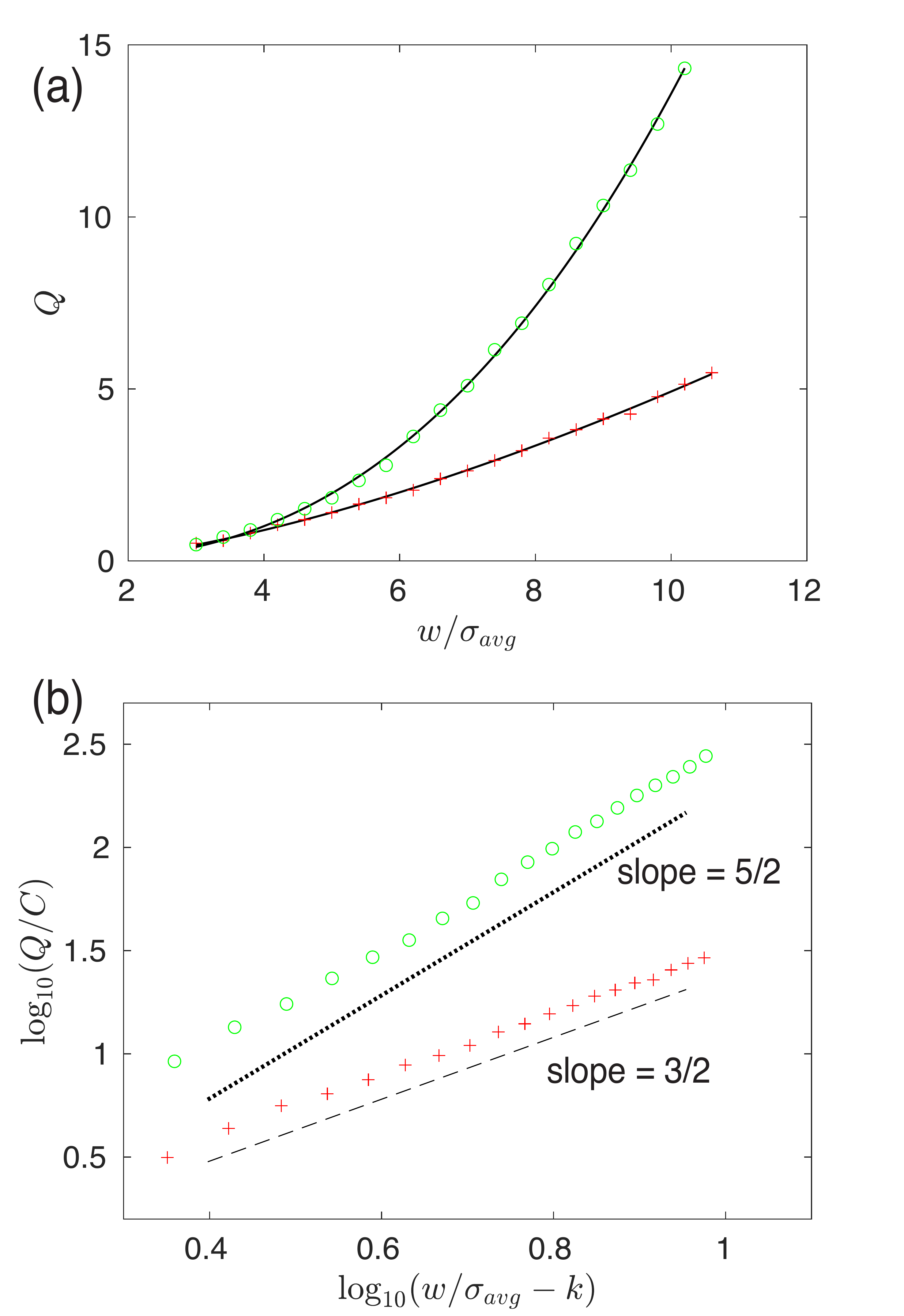}
\centering
\caption{
(a) Volume flow rate $Q$ versus orifice diameter $w/\sigma_{\rm avg}$ for hopper flows in 3D using the SP model with $E_{sp}=10^2$ and either kinetic friction only, $\mu/\mu_0 = 10\sqrt{10}$ (circles), or viscous drag only, $\zeta/\zeta_0 = 1/\sqrt{10}$ (crosses) for the dissipative forces. The solid curves in (a) are fits to the power-law scaling relation in Eq.~\ref{eq:1}. In (b), we show $\log_{10}(Q/C)$ versus $\log_{10}(w/\sigma_{\rm avg}-k)$ for the data in (a), and the dotted and dashed lines have slopes of $3/2$ and $5/2$, respectively.} 
\label{fig:3D}
\end{figure}

\section{Experimental Methods}
\label{expmethods}

Below, we will compare the simulation results for hopper flows using the SP and DP models in 2D to experimental studies of quasi-2D hopper flows of oil droplets in water. In this section, we describe the details of the experimental studies. We consider silicon oil-in-water emulsions undergoing gravity-driven hopper flows in narrow channels.

The oil-in-water emulsions are prepared through the aid of a Micronit focused-flow microfluidic device.  This device is capable of producing hundreds of droplets with volumes set by the relative flow rates between the continuous and dispersed phases \cite{Shah2008,Utada2007}.  To stabilize the emulsions, the droplets are suspended and created in a $5\%$ Tween 20 nonionic detergent solution \cite{Xin2013}.  The density of the droplets is $\rho_{\rm oil} \sim 0.936$ g/ml, and they are suspended in water with density $\rho_{\rm water} \sim 0.997$ g/ml.

The oil-in-water emulsions produced by the flow-focused microfluidic device are then injected between two $75\times50$ mm$^2$ microscope slides separated by a thin sheet of either a glass cover-slip or laser-cut plastic, ranging in thickness from $180$ to $220$ $\mu$m, in accordance with prior work on the clogging of emulsion droplets~\cite{Hong2017}. In both cases, the thickness of the sheet is sufficiently small (smaller than the smallest droplet diameter) to keep droplets from stacking. Hence, the thickness of the spacer, which also doubles as the hopper itself, confines the droplets to nearly two-dimensions \cite{Desmond2013}. The chambers are sealed with Norland Optical Adhesive 68 and placed under ultraviolet light to harden.  Inside the chambers, the droplets generally have polydispersity in size between 6-15\% (where the polydispersity is defined by the standard deviation of the droplet diameter divided by the mean).  The mean diameter of the droplets ranged from $250$-$400$~$\mu$m between different experiments. However, the mean diameter was always larger than the chamber thickness, which ensures that the droplets are quasi-two-dimensional.  Across all experiments, the average droplet diameter is $\sigma_{\rm avg}\sim 315$~$\mu$m.  Occasionally, the oil droplets coalesce into larger droplets with diameters significantly greater than 400 $\mu$m; however, these larger droplets are either among the last droplets to pass through the opening, thus acting solely as sources of pressure, or the very first, thus contributing nothing to the subsequent flow.

Two hopper geometries were used for these experimental studies:  the first has two walls oriented at $45^\circ$, and the second has one $45^\circ$ wall facing a $0^\circ$ wall (that is, a wall parallel to the direction of droplet motion).  The $45^\circ/0^\circ$ geometry is made with cover-slip glasses.  The $45^\circ/45^\circ$ geometry uses thin sheets of plastic that are laser cut into the desired shapes.  The length and width of the glass and plastic hoppers are chosen to appropriately fit the microscope slide, ensuring room for hundreds of droplets, but small enough to ensure that the droplets have enough space to clear the orifice unimpeded by droplets that had gathered outside of the hopper. The range of the hopper openings is $w/\sigma_{\rm avg} \sim 1.7$ to $12.3$. 

To initialize the experiments, a large air bubble is introduced into the sample chamber to clog the opening.  This allows droplets to stack against the bubble and create a well-packed initial condition.  We then press the sample chamber which induces the air bubble to exit, thus initiating the oil droplet flow.  To observe the flow we rotate a microscope $90^\circ$, aligning the stage parallel to the direction of gravity and viewing the sample with a horizontally directed microscope objective ($1.6\times$).  An external lamp is used for illumination and images are taken with a digital camera recording at $0.75$ fps.  Using image analysis, we obtain the droplet positions and areas, and use standard methods to track the droplet motion \cite{Crocker1996}.

Similar to the simulations, the time and area flow rate units in the experiments are defined as $t_0 = \sqrt{\sigma_{\rm avg}/g_{\rm eff}}$ and $Q_0 = \sigma_{\rm avg}^2/t_0$. We use the mean diameter across all of the experiments, $\sigma_{avg} \sim 315$~$\mu$m, and $g_{\rm eff} = g(\rho_{\rm water} - \rho_{\rm oil})/\rho_{\rm oil}$ is the acceleration imposed by oil-in-water buoyancy and $g \sim 9.8$ m/s$^2$ is the gravitational constant.

\section{Results}
\label{results}

In this section, we describe the results of our numerical simulations of hopper flows of the soft particle model (SP) in 2D and 3D and the deformable particle (DP) model in 2D. We investigate the scaling of the flow rate $Q$ versus the orifice width $w$ as a function of the ratio of the viscous drag and kinetic friction coefficients, particle deformability, surface roughness, and spatial dimension. We find that the flow rate $Q$ scales as a power-law in the orifice width $w/\sigma_{\rm avg}$ with a cutoff $k$, $Q = C(w/\sigma_{\rm avg}-k)^{\beta}$. The power-law scaling exponent $\beta$ depends strongly on the ratio of the viscous drag and kinetic friction coefficients $\lambda =\zeta/\mu$, but it does not depend on the particle deformability or surface roughness. In particular, if the particles only experience kinetic friction, without viscous drag, $\beta=d-1/2$, as found by Beverloo and others for hopper flows of granular materials. However, if the particles only experience viscous drag, without kinetic friction, $\beta=d-3/2$. Further, we show that the scaling exponent $\beta$ varies continuously with $\lambda$ between $\beta = d-1/2$ in the $\lambda \rightarrow 0$ limit and $d-3/2$ in the $\lambda \rightarrow \infty$ limit, with a midpoint $\lambda_c$ that decreases with decreasing hopper opening angle $\theta_w$. We show that the change in the power-law scaling exponent $\beta$ is associated with changes in the spatio-temporal dynamics of the flows. In particular, the gradient in the velocity profile varies more strongly with the orifice size $w$ for flows with $\beta=d-3/2$ than those with $\beta=d-1/2$. We then show that the offset $k$ at which $Q(k\sigma_s)=0$ decreases from values above $3$ to below $1$ as the particle deformability increases. We also find that both the soft and deformable particle models in the $\lambda \rightarrow \infty$ limit are able to recapitulate $Q(w)$ obtained from experimental studies of quasi-2D hopper flows of oil droplets in water.
\begin{figure}[!ht]
\includegraphics[width=0.95\columnwidth]{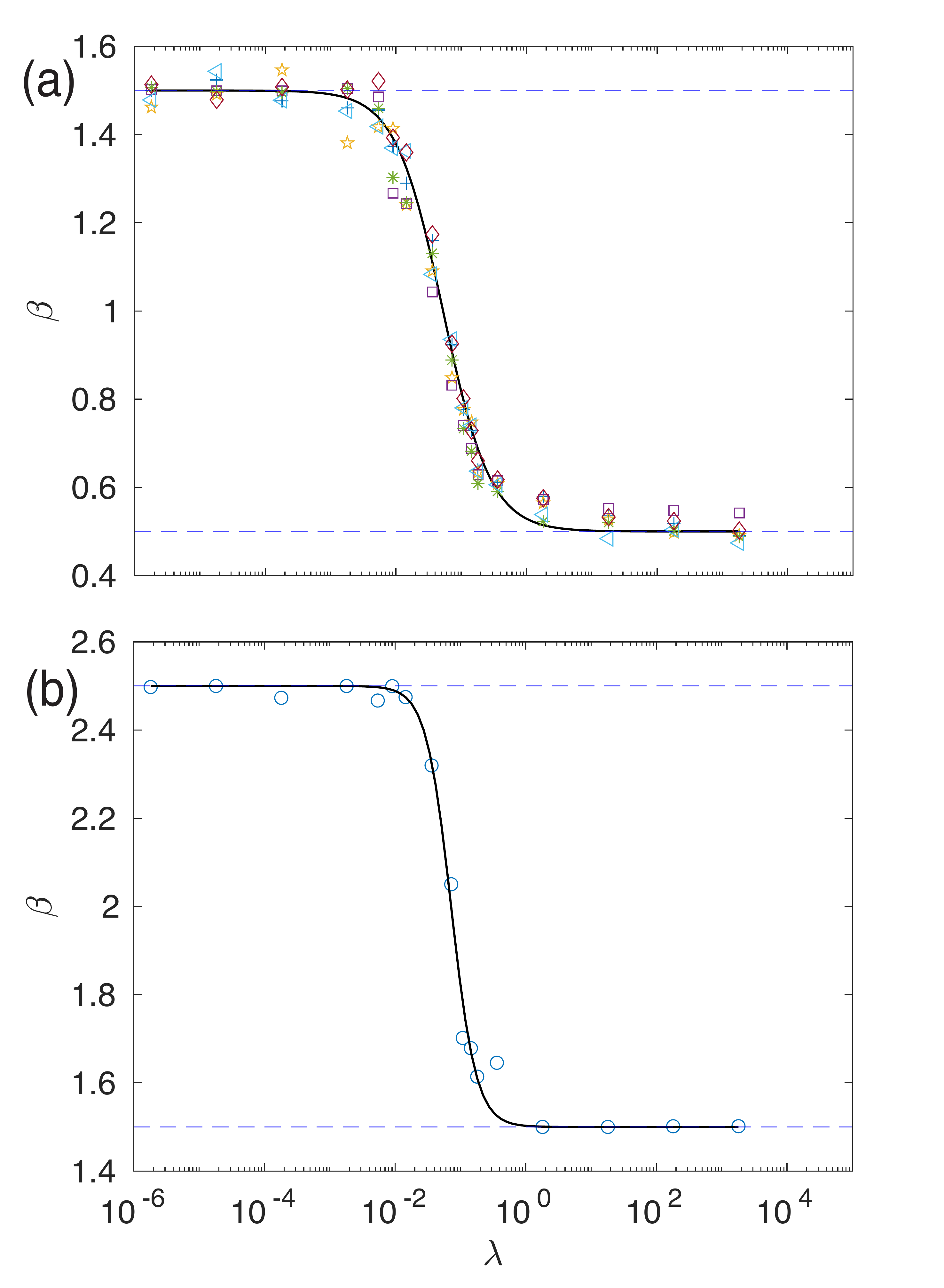}
\centering
\caption{Power-law scaling exponent $\beta$ from Eq.~\ref{eq:1} plotted versus the ratio of the viscous drag and kinetic friction coefficients $\lambda=\zeta/\mu$ for hopper flows in (a) 2D and (b) 3D. In 2D, we consider the SP model with $E_{sp}=50$ (triangles) and $10^3$ (diamonds), frictional DP model with  $K_l=10$ and $K_b=10^{-1}$ (stars), and the frictionless DP model with $K_l=10$ and $K_b=10^{-1}$ (crosses), $K_l=10$ and $K_b=10^2$ (squares), and $K_l=10^3$ and $K_b=10^{-1}$ (asterisks), all with hopper opening angle $\theta_w = 90^{\circ}$. In 3D, we consider the SP model with $E_{sp}=10^2$ (circles). In (a) and (b), the solid curves are fits to the sigmoid in Eq.~\ref{eq:b_sig} and the horizontal dashed lines indicate $\beta = 5/2$, $3/2$, and $1/2$.} 
    \label{fig:drag2friction}
\end{figure}

In Fig.~\ref{fig:2} (a), we show the area flow rate $Q$ versus the orifice width $w/\sigma_{\rm avg}$ for the soft particle and deformable particle models with kinetic friction only (i.e. $\mu/\mu_0 =10\sqrt{10}$ and $\zeta=0$) in 2D for hopper opening angle $\theta_w =90^{\circ}$.  We compare results for the SP model with $E_{sp} = 10^2$ and $10^4$, the frictionless DP model with $K_l=10$ and $K_b = 10^{-1}$, $K_l=10$ and $K_b =10$, and $K_l=10^3$ and $K_b =10^{-1}$, and the frictional DP model with $K_l =10$ and $K_b=10$. $Q(w)$ for all of these systems can be fit to the power-law scaling relation in Eq.~\ref{eq:1}. While $C$ and $k$ for these systems vary, the power-law scaling exponent $\beta=3/2$ is the same for all models as shown in Fig.~\ref{fig:2} (b). 
(Note that both the SP and DP models can be studied in the rigid-particle limit, i.e. $E_{sp} \rightarrow \infty$ for the SP model and $K_b \rightarrow \infty$ for the DP model. In this limit, $Q(w)$ is the same for both models as shown in Appendix B.)  Since we compared systems with different values of $K_{sp}$, $K_b$, and $K_l$ and with different values of surface roughness and obtained the same values of $\beta$, these results emphasize that $\beta$ does not depend strongly on particle deformability and surface roughness. 

\begin{figure}[!ht]
\includegraphics[width=0.95\columnwidth]{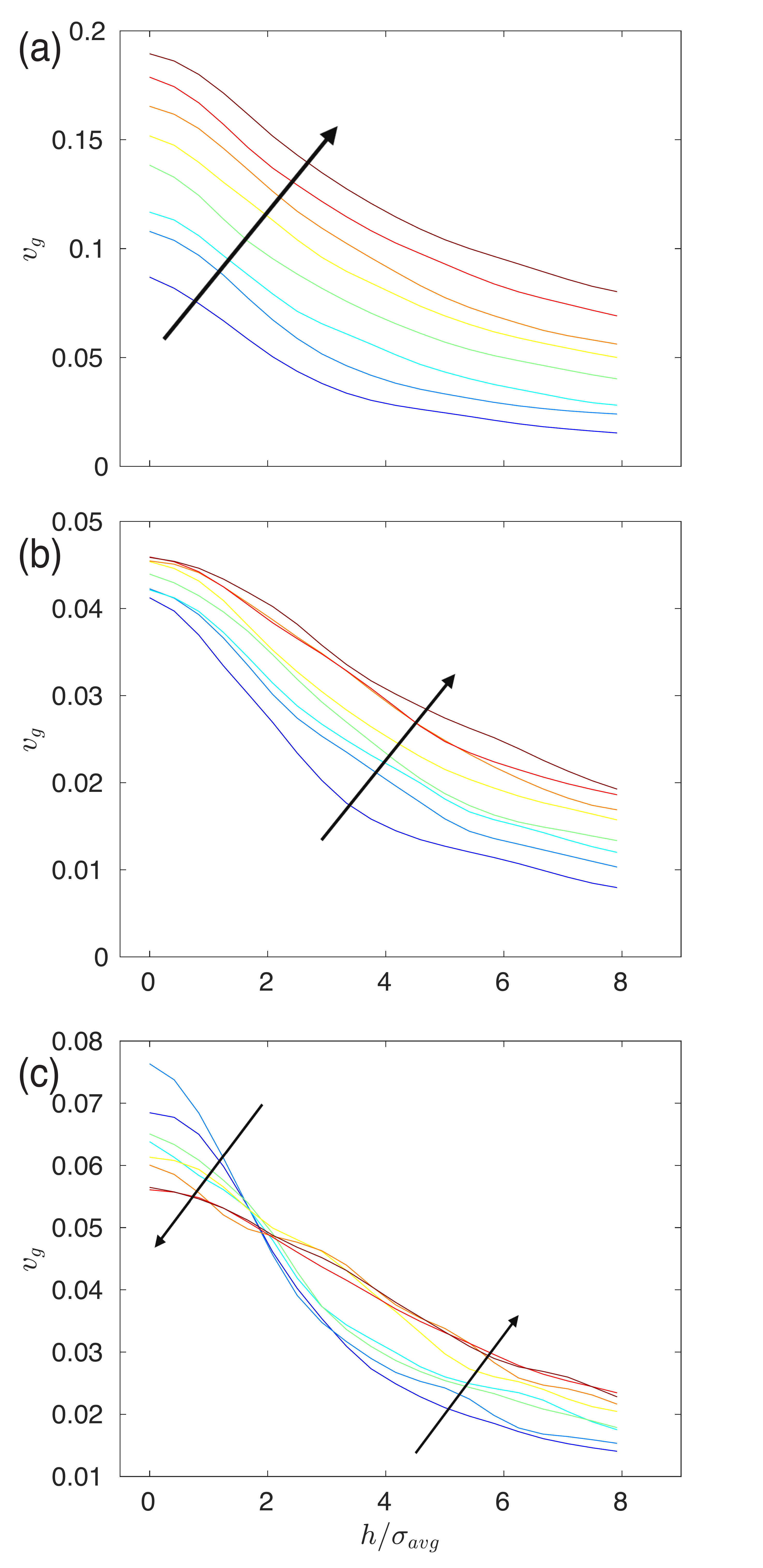}
\centering
\caption{Average speed of the particles in the direction of gravity at the center of the hopper $v_g$ as a function of distance $h/\sigma_{\rm avg}$ above the hopper orifice in 2D using the SP model with $E_{sp} = 10^2$ and dissipative forces (a) with $\lambda \rightarrow 0$ that yield $\beta =d-1/2$, (b) with $\lambda \sim \lambda_c$ that yield $\beta \sim d-1$, and (c) with $\lambda \rightarrow \infty$ that yield $\beta =d-3/2$. The arrows indicate increasing orifice diameters from $w/\sigma_{avg} = 4.0$ (blue) to $10.6$ (red).} 
    \label{fig:vprofile}
\end{figure}

In Fig.~\ref{fig:2} (c), we show similar results for $Q$ versus $w/\sigma_{\rm avg}$ for same 2D models, but for systems with viscous drag forces only (i.e. $\mu =0$ and $\zeta/\zeta_0=1/\sqrt{10}$) for the dissipative forces. All of the data can also be fit to the power-law scaling relation in Eq.~\ref{eq:1}. Again, $C$ and $k$ vary, but the power-law scaling exponent $\beta=1/2$ is the same for all models, as shown in Fig.~\ref{fig:2} (d).  Clearly, the power-law scaling exponent $\beta$ does not depend on particle deformability and surface roughness, but it depends strongly on the type of dissipative forces that are included.  

In Fig.~\ref{fig:3D} (a), we show similar results for the volume flow rate $Q$ versus orifice width $w/\sigma_{\rm avg}$ for the SP model in 3D 
with $E_{sp}=10^2$ and either kinetic friction forces only ($\mu/\mu_0=10\sqrt{10}$, $\zeta=0$) or viscous drag forces only ($\mu=0$, $\zeta/\zeta_0=1/\sqrt{10}$).  $Q(w)$ for both systems can be fit by Eq.~\ref{eq:1} and have power-law scaling exponents $\beta=5/2$ and $3/2$ in the limits $\lambda \rightarrow 0$ and $\infty$, respectively, as shown in Fig.~\ref{fig:3D} (b). Figs.~\ref{fig:2} and~\ref{fig:3D} illustrate that $\beta=d-1/2$ in the $\lambda \rightarrow 0$ limit and $\beta =d-3/2$ in the $\lambda \rightarrow \infty$ limit. 

\begin{figure}[!ht]
\includegraphics[width=0.9\columnwidth]{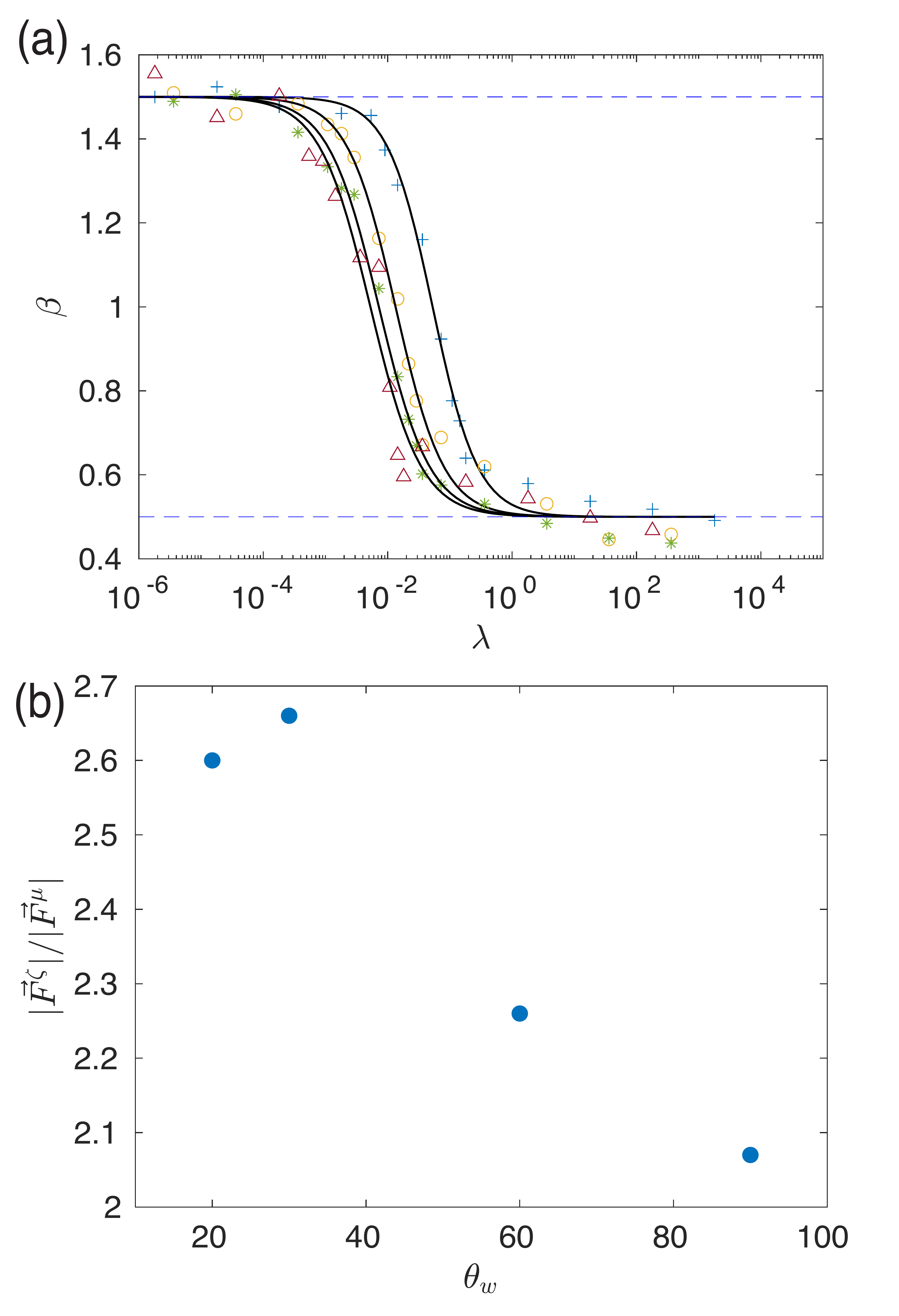}
\centering
\caption{(a) Power-law scaling exponent $\beta$ versus the ratio $\lambda$ of the viscous drag and kinetic friction coefficients for the frictionless DP model in 2D with $K_{l} = 10$ and $K_b = 10^{-1}$ for hopper opening angles $\theta_w = 90^\circ$ (crosses), $60^\circ$ (circles), $30^\circ$ (asterisks), and $20^\circ$ (triangles). The solid lines are fits to Eq.~\ref{eq:b_sig}. The horizontal dotted and dashed lines indicate $\beta=3/2$ and $1/2$. (b) Ratio of the average magnitude of the drag force on a particle to the average magnitude of the kinetic friction force on a particle $|{\vec F}^{\zeta}|/|{\vec F}^{\mu}|$ plotted versus $\theta_w$ for the systems in (a) at $\lambda = 10^{-2}$.} 
    \label{fig:hopperAngle}
\end{figure}

What is the value of the power-law exponent $\beta$ at intermediate values of $\lambda$?  In Fig.~\ref{fig:drag2friction}, we show $\beta$ from fits of $Q(w)$ to Eq.~\ref{eq:1} for the SP and DP models in 2D (for hopper opening angle $\theta_w = 90^{\circ}$) and the SP model in 3D versus the ratio of the viscous drag and kinetic friction coefficients $\lambda$. $\beta$ varies continuously with $\lambda$ in both 2D and 3D and can be described by a sigmoidal function:
\begin{equation}
\beta = \frac{1}{2}\left(d-\tanh\left[\log_{10}\left(\lambda-\lambda_c\right)^{1/b}\right]\right),
\label{eq:b_sig}
\end{equation}
where $\lambda_c \sim 0.05$ and $\sim 0.07$ in 2D and 3D is the characteristic value at which the power-law scaling exponent reaches the midpoint $\beta_c = d-1$ and $0< 1/b < 1$ is the stretching exponent. We show explicitly in 2D that $\beta(\lambda)$ does not depend on particle deformability and surface roughness. Further, these results do not depend on the number of particles $N>800$ as shown in Appendix C. We find similar results for $\beta(\lambda)$in 3D. 

\begin{figure}[!ht]
\includegraphics[width=0.95\columnwidth]{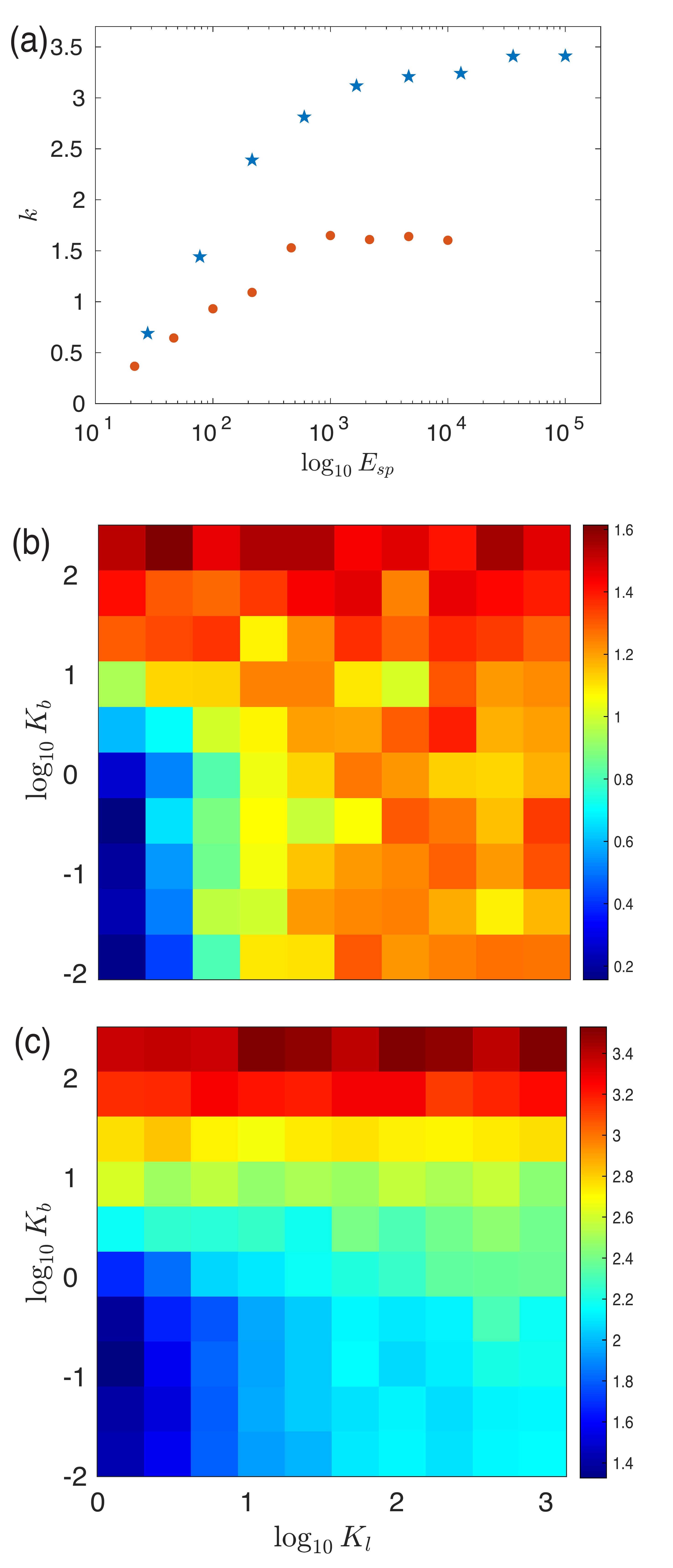}
\caption{(a) The offset $k$ obtained from fits of the area flow rate $Q(w)$ to Eq.~\ref{eq:1} versus $E_{sp}$ (with $\theta_w =90^{\circ}$) using the 2D SP model in the $\lambda \rightarrow \infty$ limit (stars) and $\lambda \rightarrow 0$ limit (circles). The offset $k$ for the 2D frictionless DP model on a color scale (b) from $0$ (blue) to $1.7$ (red) in the $\lambda \rightarrow 0$ limit (with $\theta_w =90^{\circ}$) and (c) from $1$ (blue) to $3.5$ (red) in the $\lambda \rightarrow \infty$ limit (with $\theta_w =90^{\circ}$) as a function of the perimeter $K_l$ and bending $K_b$ energy scales.} 
\label{fig:heatMap}
\end{figure}

What is different about the spatiotemporal dynamics of the hopper flows with different values of the power-law scaling exponent $\beta$? To address this question, we calculate the velocity profiles in systems with different values of $\beta$. In Fig.~\ref{fig:vprofile}, we show (for the SP model in 2D with $\theta_w = 90^{\circ}$) the average speed of the  particles in the direction of gravity at the center of the hopper $v_g$ as a function of the distance above the hopper orifice $h/\sigma_{\rm avg}$ for three ratios of the dissipative forces, $\lambda \rightarrow 0$, $\lambda \sim \lambda_c$, and $\lambda \rightarrow \infty$.  
To smooth the velocity profile, we define $v_g$ at location $\vec{r}$ as $v_g(\vec{r}) = \sum_{i=1}^{N} v_{gi} \phi({\vec r}-\vec{r}_i)$, where $\vec{r}_i$ and $v_{gi}$ are the position and speed in the direction of gravity of particle $i$ and $ \phi(\vec{r}-\vec{r}_i) =(\sqrt{2 \pi} \sigma_{\rm avg})^{-2} \exp(-|\vec{r}-\vec{r}_i|^2/2\sigma_{\rm avg} ^2)$ is a Gaussian coarse-graining function\cite{PhysRevLett.114.238002,Goldhirsch2010}.
For systems with $\lambda \rightarrow 0$ and $\beta = d - 1/2$, $v_g(h=0) \sim w^\beta/w^{d-1} \sim w^{1/2}$, and thus $v_g(h=0)$ increases with the orifice diameter $w$, as shown in Fig.~\ref{fig:vprofile} (a). For systems with $\lambda \rightarrow \infty$ and $\beta = d - 3/2$, $v_g(h=0) \sim w^\beta/w^{d-1} \sim w^{-1/2}$, and thus $v_g(h=0)$ decreases with increasing $w$, as shown in Fig.~\ref{fig:vprofile} (c). In contrast, the average speed in the direction of gravity far from the hopper orifice, $v_g(h \rightarrow \infty) \sim w^{\beta}/W\sim w^{\beta}$, increases with $w$ for all values of $\lambda$, as shown in Fig.~\ref{fig:vprofile} (a)-(c). Because of the difference in how $v_g(0)$ and $v_g(\infty)$ depend on the orifice width $w$ for different values of $\lambda$, the gradient of the velocity profile $dv_g/dh$ can easily distinguish flows with small versus large values of $\lambda$. As shown in Fig.~\ref{fig:vprofile} (a), for $\lambda \rightarrow 0$, $dv_g/dh$ does not depend strongly on $w$, suggesting a weak variation of the pressure profile on $w$. However, for $\lambda \rightarrow \infty$, $dv_g/dh$ decreases strongly with increasing $w$, indicating large pressure profiles in systems with small $w$.  In this limit, the large differences in viscous drag forces caused by the velocity difference $v(0)-v(\infty)$ are balanced by overlap forces, which give rise to the large pressure profile.  As expected, $v_g$ near the orifice for the intermediate case $\lambda \sim \lambda_c$ possess very weak dependence on $w$.  

\begin{figure}[!ht]
\centering
\includegraphics[width=0.9\columnwidth]{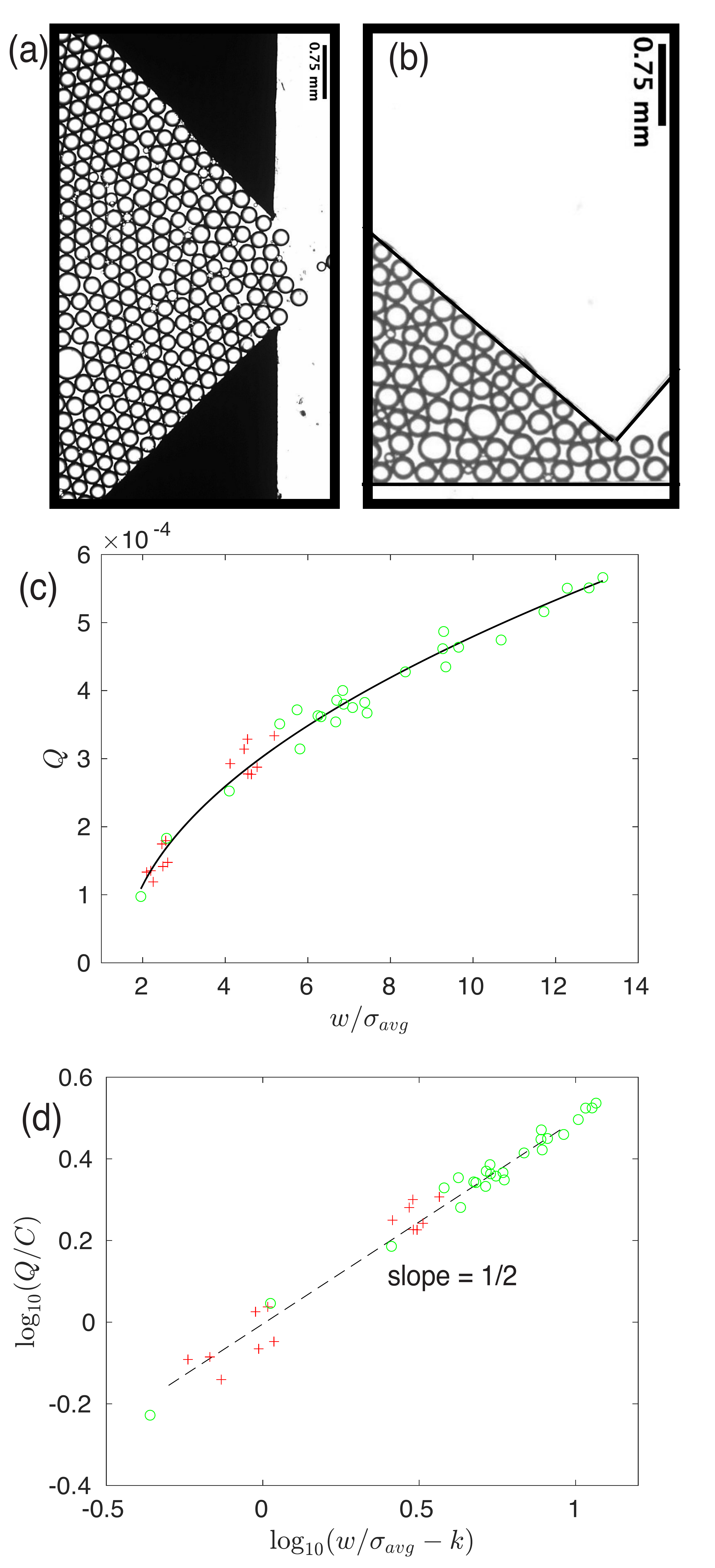}
\caption{Oil-in-water emulsions flowing through a plastic hopper with orifice diameter $w \sim 180$ $\mu$m and (a) $45^{\circ}$/$45^{\circ}$ and (b) $45^{\circ}$/$0^{\circ}$ wall geometries. The droplets have an average diameter $\sigma_{\rm avg} \sim 304$ $\mu$m with a polydispersity $\Delta \sigma/\sigma_{\rm avg} \sim 7 \%$. (c) Area flow rate $Q$ plotted versus orifice diameter $w$. The solid line provides a fit to Eq.~\ref{eq:1} with $\beta \sim 0.49$, $k \sim 1.47$, and $C \sim 1.6 \times 10^{-4}$. (d) $Q/C$ plotted versus $\log_{10} (w/\sigma_{\rm avg}-k)$ for the data in (c). The dashed line has a slope of $1/2$. In (c) and (d), we show data for both $45^{\circ}$/$45^{\circ}$ (circles) and $45^{\circ}$/$0^{\circ}$ (crosses) wall geometries.}
\label{fig:exp}
\end{figure}

We have shown that $\beta(\lambda)$ does not depend on particle deformability and surface roughness, but it does depend strongly on the nature of the dissipative forces (i.e. whether viscous drag or kinetic frictional forces dominate) and the resulting velocity profile in the hopper. These results suggest that $\beta(\lambda)$ can be altered by varying the hopper opening angle $\theta_w$ since changes in $\theta_w$ modify the velocity profile. In Fig.\ref{fig:hopperAngle} (a), we show $\beta(\lambda)$ from hopper flows using the frictionless DP model in 2D with $\theta_w = 90^{\circ}$, $60^{\circ}$, $30^{\circ}$, and $20^{\circ}$.  Over this range in $\theta_w$, the characteristic $\lambda_c$ at which $\beta$ reaches its midpoint decreases from $5 \times 10^{-2}$ to $6 \times 10^{-3}$. As the hopper wall angle $\theta_w$ decreases (i.e. the hopper walls become more aligned with the direction of gravity), in the regime $\lambda \sim \lambda_c$, the ratio of the  
average force stemming from the viscous drag to that stemming from the kinetic friction $|\vec{F}^\zeta|/|\vec{F}^\mu|$ increases (Fig.\ref{fig:hopperAngle} (b)), and thus $\lambda_c$ must decrease with decreasing $\theta_w$. In the low-$\theta_w$ limit (i.e. $\theta_w \le 30^{\circ}$), the ratio stops increasing and $\lambda_c$ reaches a plateau value, $\sim 5 \times 10^{-3}$.  Note that the time required to reach steady-state diverges as $\theta_w \rightarrow 0$, and thus we are limited in the values of $\theta_w$ that we can study. 

In contrast to the power-law scaling exponent $\beta$, the offset $k$ at which $Q(k\sigma_{\rm avg})=0$ depends on particle deformability and surface roughness. 
Previous studies have shown that $k$ varies from $\sim 1.3$ to $2.9$ as the static friction coefficient increases\cite{BEVERLOO1961260,ANAND20085821}.
In Fig.~\ref{fig:2}, we show similar results that $k$ increases with surface roughness for the DP model. How does the offset $k$ depend on particle deformability? In Fig.~\ref{fig:heatMap} (c), we show 
$k$ as a function of the perimeter $K_l$ and bending $K_b$ energy scales for the frictionless DP model in 2D in the $\lambda \rightarrow \infty$ limit (for $\theta_w = 90^{\circ}$). At small $K_l$, $k$ increases from $\sim 1$ to $\sim 3.5$ as $K_b$ increases from $10^{-2}$ to above $10^2$, suggesting the formation of transient multi-particle arches in the rigid-particle limit. (Note that we have shown in Appendix B that the DP model reaches the hard-particle limit for $K_b > 10^{2}$.) We find similar results for the 2D SP model for $\lambda \rightarrow \infty$ (Fig.~\ref{fig:heatMap} (a)): the offset $k$ increases from $k \sim 1$ to $3.5$ as $E_{sp}$ approaches the rigid-particle limit. At small $K_b$ for the DP model, $k$ increases, but only from $k \sim 1$ to $2$ as $K_l$ increases from $1$ to $10^3$, suggesting the formation of small arches and increased particle rigidity. However, $K_l \gg 10^3$ is required to reach $k \sim 3.5$ as found in the rigid-particle limit for the DP model when increasing the bending energy.  In addition, we find that the prefactor in Eq.~\ref{eq:1} (with $N=1600$) $C \sim 0.45$ for all $K_b$ and $K_l$ for the DP model and all $E_{sp}$ for the SP model, emphasizing that $C$ is weakly dependent on particle deformability. We show the system size dependence of the prefactor in Appendix C. 

For hopper flows with $\lambda \rightarrow 0$, the increase in the offset $k$ is much less pronounced. (See Fig.~\ref{fig:heatMap} (a) and (b).) For example, for the 2D DP model, $k < 1$ for $K_b = 10^{-2}$ and $k \sim 1.5$ for $K_b = 10^2$ in the rigid-particle limit.  Thus, large multi-particle arches do not form frequently in $\lambda \rightarrow 0$ hopper flows. Again, $C \sim 0.15$ for all $K_b$, $K_l$ and $E_{sp}$ values (for $N=1600$).

As discussed in the Introduction, numerous experimental studies have shown that hopper flows of granular materials with static and kinetic frictional forces posses $\beta =d-1/2$ and can be modeled quantitatively using the SP model.  Here, we present the results from quasi-2D experiments of hopper flows of oil droplets in water. (See Fig.~\ref{fig:exp} (a) and (b).)  Unlike the simulations where the number of particles in the hopper is kept constant (by replenishing them when particles exit), the number of particles in the hopper experiments decreases with time.  The hopper flow is driven by hydrostatic pressure, which scales with the height $h_{\rm max}$ of the droplet pile pushing out of the opening.  Given the triangular geometry, this distance can be related to the number of droplets $N$ that have yet to exit, $h_{\rm max} \sim \sqrt{N}$.  Hence, the droplet flux can be written as  
\begin{equation}\label{first}
    \frac{dN}{dt}=c_0\sqrt{N},
\end{equation}
where $c_0$ has units of inverse time.  This relation is experimentally observed for a large range of $N$, with slight deviations as the first $\sim 100$ and last $\sim 100$ droplets flow out due to transient effects.  Fitting the steady-state data gives values for $c_0$, which we non-dimensionalize as $Q = c_0 \sqrt{\sigma_{\rm avg}/g_{\rm eff}}$. Fig.~\ref{fig:exp} (c) and (d) show the results for the area flow rate $Q$ versus the orifice width $w/\sigma_{\rm avg}$. While we have two experimental geometries, the results for the area flow rate $Q$ are identical.  $Q(w)$ can be fit by the power-law scaling relation in Eq.~\ref{eq:1} with $\beta \sim 0.49$ and $k\sim 1.5$. These values of $\beta$ and $k$ are consistent with the simulation results for $\lambda \rightarrow \infty$ and $E_{sp} \sim 10^2$ for the SP model and the $k \sim 1.5$ contour for the frictionless DP model in Fig.~\ref{fig:heatMap} (c).  These results emphasize that the kinetic frictional forces are weak relative to the viscous drag forces in hopper flows of oil droplets in water.

\begin{figure}[!ht]
\includegraphics[width=0.9\columnwidth]{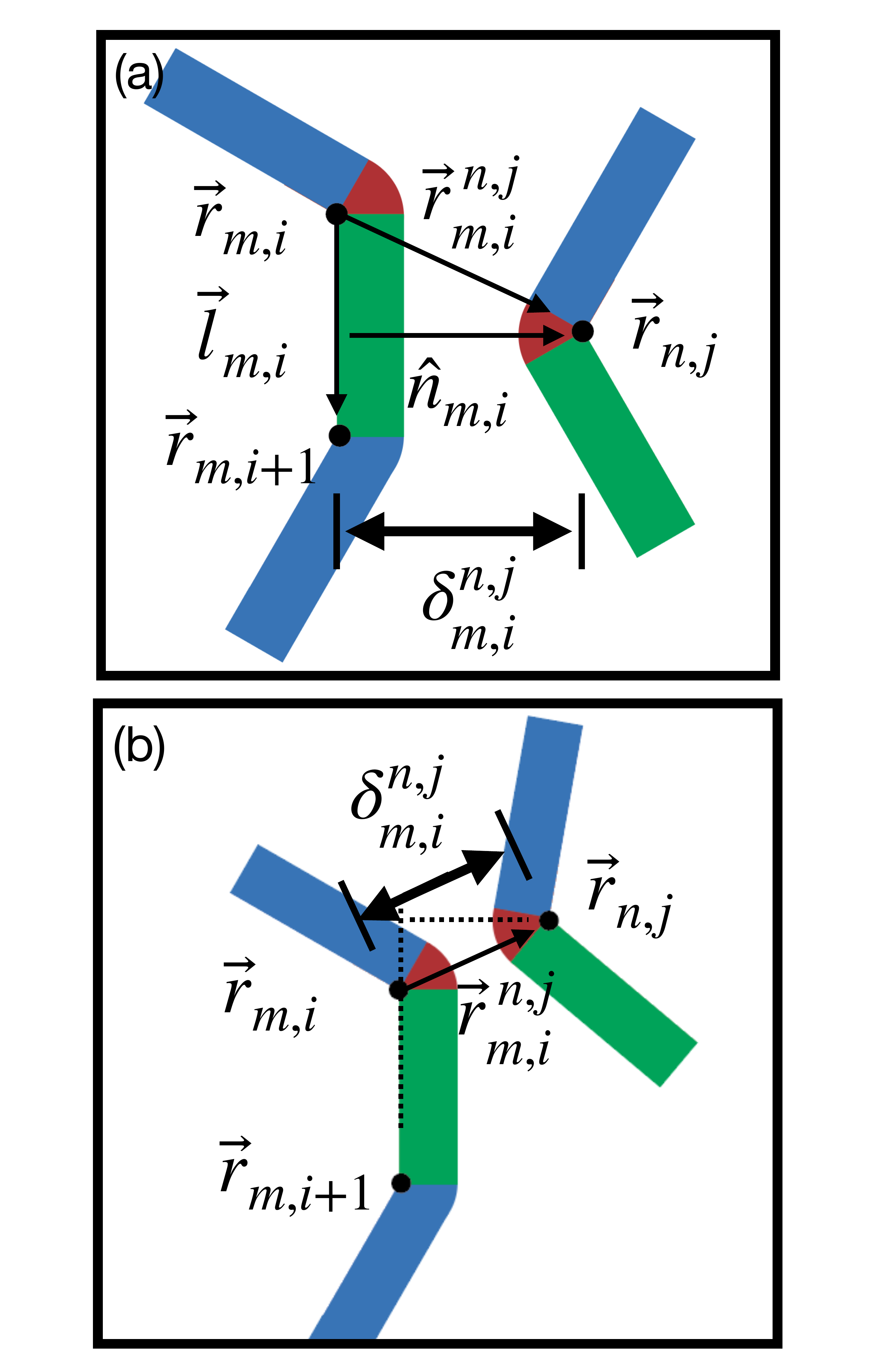}
\centering
\caption{Schematic of the frictionless DP model to illustrate the contact distance $\delta_{m,i}^{n,j}$ between vertex $i$ on particle $m$ with position ${\vec r}_{m,i}$ and vertex $j$ on particle $n$ with position ${\vec r}_{n,j}$.  ${\vec l}_{m,i}$ is the vector pointing from ${\vec r}_{m,i}$ to ${\vec r}_{m,i+1}$ and ${\hat n}_{m,i} \cdot {\vec l}_{m,i} =0$. The definition of $\delta_{m,i}^{n,j}$ depends on the location of the intersection point $P$ of the line along ${\vec l}_{m,i}$ and the line that is perpendicular to ${\vec l}_{m,i}$ that includes ${\vec r}_{n,j}$.  If point $P$ is between ${\vec r}_{m,i}$ and ${\vec r}_{m,i+1}$, $\delta_{m,i}^{n,j} = {\vec r}_{m,i}^{~n,j} \cdot {\hat n}_{m,i}$ as shown in (a), otherwise $\delta_{m,i}^{n,j} = |{\vec r}_{m,i}^{~n,j}|$ as shown in (b).} 
\label{fig:frictionless}
\end{figure}

\section{Discussion and Conclusions}
\label{discussion}

In this article, we carried out extensive numerical simulations of gravity-driven hopper flows of particulate media in 2D and 3D using the soft (SP) and deformable particle (DP) models.  We found several important results.  First, we showed quite generally that the flow rate $Q$ versus orifice width $w$ obeys the power-law scaling relation: $Q(w) = C(w/\sigma_{\rm avg}-k)^{\beta}$. While $k$ depend on the particle deformability and surface roughness, the exponent $\beta$ does not.  Instead, $\beta$ is controlled by the ratio of the viscous drag to the kinetic friction coefficients $\lambda$ and $\beta$ varies continuously from $\beta = d-1/2$ in the $\lambda \rightarrow 0$ limit to $\beta = d-3/2$ in the $\lambda \rightarrow \infty$ limit. The midpoint $\beta_c(\lambda_c)$ can be tuned by varying the hopper opening angle $\theta_w$ since it can alter the ratio of the average viscous drag force to the average kinetic friction force. The spatiotemporal dynamics of the flows differ for systems with different power-law exponents. In particular, the gradients of the velocity and pressure profiles vary more strongly with the orifice width for $\beta = d-3/2$ than those with $\beta=d-1/2$. We also found the offset $k$ increases with particle stiffness until $k \sim k_{\rm max}$ in the hard-particle limit, where $k_{\rm max} \sim 3.5$ in $\lambda \rightarrow \infty$ limit and $k_{\rm max} \sim 1.6$ in $\lambda \rightarrow 0$ limit. In addition, we showed that both the SP and DP models are able to recapitulate the flow rate $Q(w)$ from experimental studies of quasi-2D hopper flows of oil droplets in water.   

These results suggest a number of promising future research directions. First, the current studies focused on nearly spherical particles. How does the power-law scaling exponent $\beta(\lambda)$ depend on particle shape?  By changing the reference shape parameter ${\cal A}_0$ in the DP model, we can determine $\beta(\lambda)$ as a function of the particle shape. Second, in the current studies, we included viscous drag and kinetic friction forces between particle pairs, but we did not include kinetic friction forces between the particles and the side walls with kinetic friction coefficient $\nu$. How does the power-law scaling exponent vary with the dimensionless ratios $\zeta/\nu$ and $\mu/\nu$ that quantify the dominant dissipative forces?  Third, in the current studies, we found that both the SP and DP models are able to recapitulate $Q(w)$ in the experimental studies of hopper flows of emulsion droplets.  However, in future studies, we seek a more quantitative approach where the simulations can recover the particle shapes during the hopper flows in experiments. To do this, we will refine the model for surface tension in the DP model.  In addition, we will simulate hopper flows of emulsion droplets in the intermittent and clogging regime for $w \sim k\sigma_{\rm avg}$.  In this regime, we expect qualitatively different results for the SP and DP models, since truly deformable particles can significantly change their shapes, but maintain their volume, to alleviate clogs in hopper flows.

\begin{figure}[!ht]
\includegraphics[width=\columnwidth]{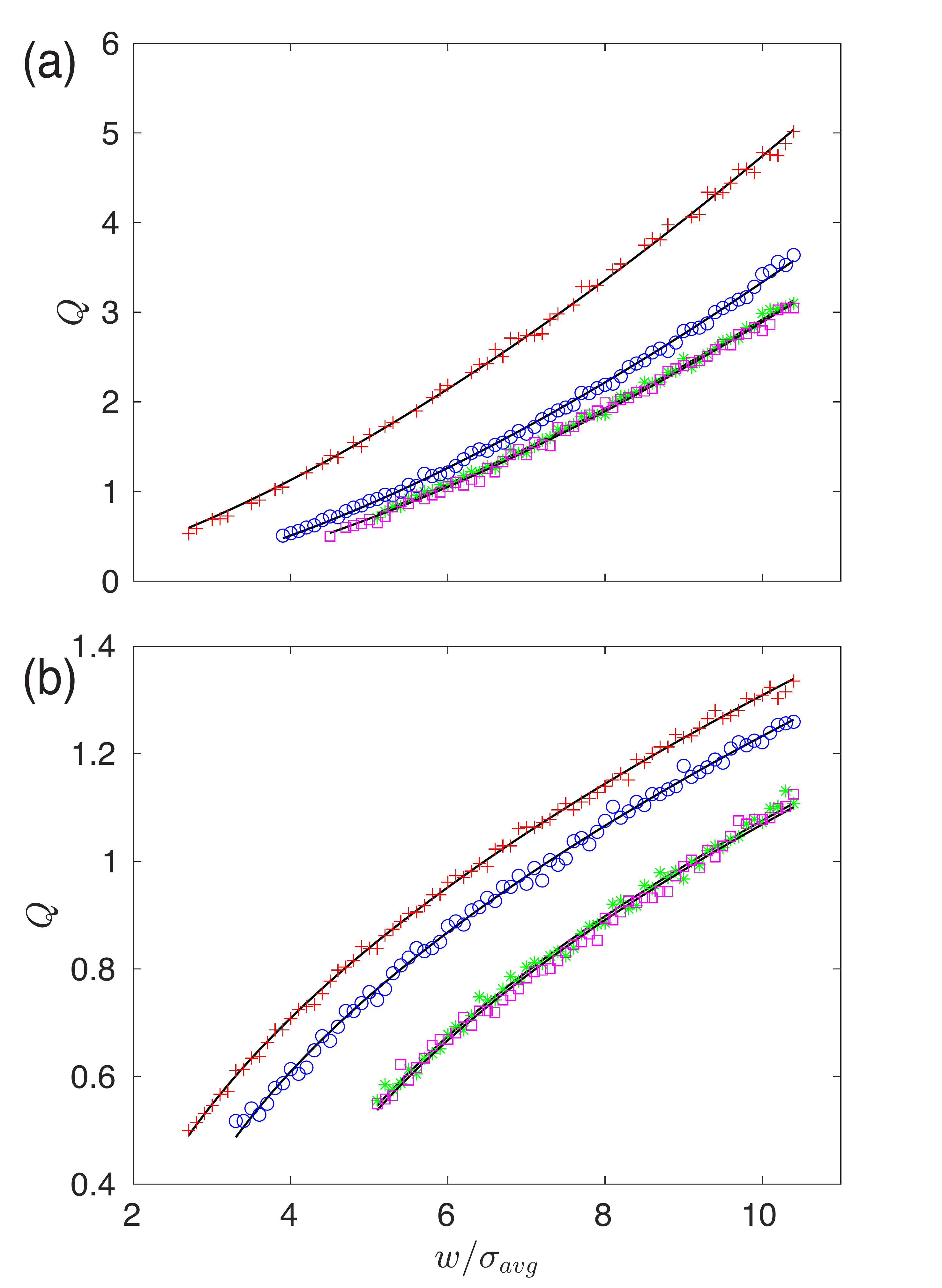}
\caption{Area flow rate $Q$ versus orifice width $w/\sigma_{\rm avg}$ for hopper flows in 2D using the SP model  $E_{sp}=10^2$ (circles) and $10^5$ (squares) and the frictionless DP model with $K_l=10$ and $K_b=10^{-1}$ (crosses) and $K_l=10$ and $K_b=10^2$ (asterisks) with (a) kinetic friction forces only ($\mu/\mu_0 = 10\sqrt{10}$, $\zeta=0$) and (b) viscous drag forces only ($\zeta/\zeta_0 = 1/\sqrt{10}$, $\mu=0$). The solid curves are fits to the power-law scaling relation for $Q(w)$ in Eq.~\ref{eq:1}. In the hard-particle limit, we find (a) $C \approx 0.12$ and $k \approx 1.6$ for $\lambda \rightarrow 0$ and (b) $C \approx 0.42$ and $k \approx 3.4$ for $\lambda \rightarrow \infty$.} 
\label{fig:rigid}
\end{figure}

\section*{Appendix A}
\label{app:A}

In this Appendix, we include more details concerning the detection of contacts between frictionless deformable particles in 2D. For frictionless deformable particles, the $i$th vertex on a given particle $m$ is modeled as a circulo-line made up of a rectangular region with length $l_{m,i}$ plus a pair of half-circular end caps with radius $\delta_m$
. Here, we describe how to calculate the closest distance $\delta_{m,i}^{n,j}$ between vertex $i$ on particle $m$ and vertex $j$ on particle $n$ as shown in Fig.\ref{fig:frictionless}. We first find the line $L$ that includes the point ${\vec r}_{n,j}$ and is perpendicular to ${\vec l}_{m,i}$. If line $L$ intersects the line along ${\vec l}_{m,i}$ at a point between ${\vec r}_{m,i}$ and ${\vec r}_{m,i+1}$, the closest distance between vertices $i$ and $j$ is the distance between ${\vec r}_{n,j}$ and the line along ${\vec l}_{m,i}$, i.e. $\delta_{m,i}^{n,j} = \vec{r}_{m,i}^{~n,j} \cdot \hat{n}_{m,i}$ as shown in Fig.\ref{fig:frictionless} (a). In this case, the repulsive pair force from $U_{\rm int}$ is in the direction of $\hat{n}_{m,i}$ (perpendicular to the surface of particle $m$), and therefore it is a frictionless interaction\cite{Kim2022}. 

If line $L$ does not intersect the line along ${\vec l}_{m,i}$ at a point between ${\vec r}_{m,i}$ and ${\vec r}_{m,i+1}$, the closest distance between vertices $i$ and $j$ is $\delta_{m,i}^{n,j} = |{\vec r}_{m,i}^{~n,j}|$ as shown in Fig.\ref{fig:frictionless} (b).  Again, in this case, the gradient of $U_{\rm int}$ is along ${\hat r}_{m,i}^{~n,j}$, and thus the repulsive interaction force is frictionless.

\begin{figure}[!ht]
\includegraphics[width=\columnwidth]{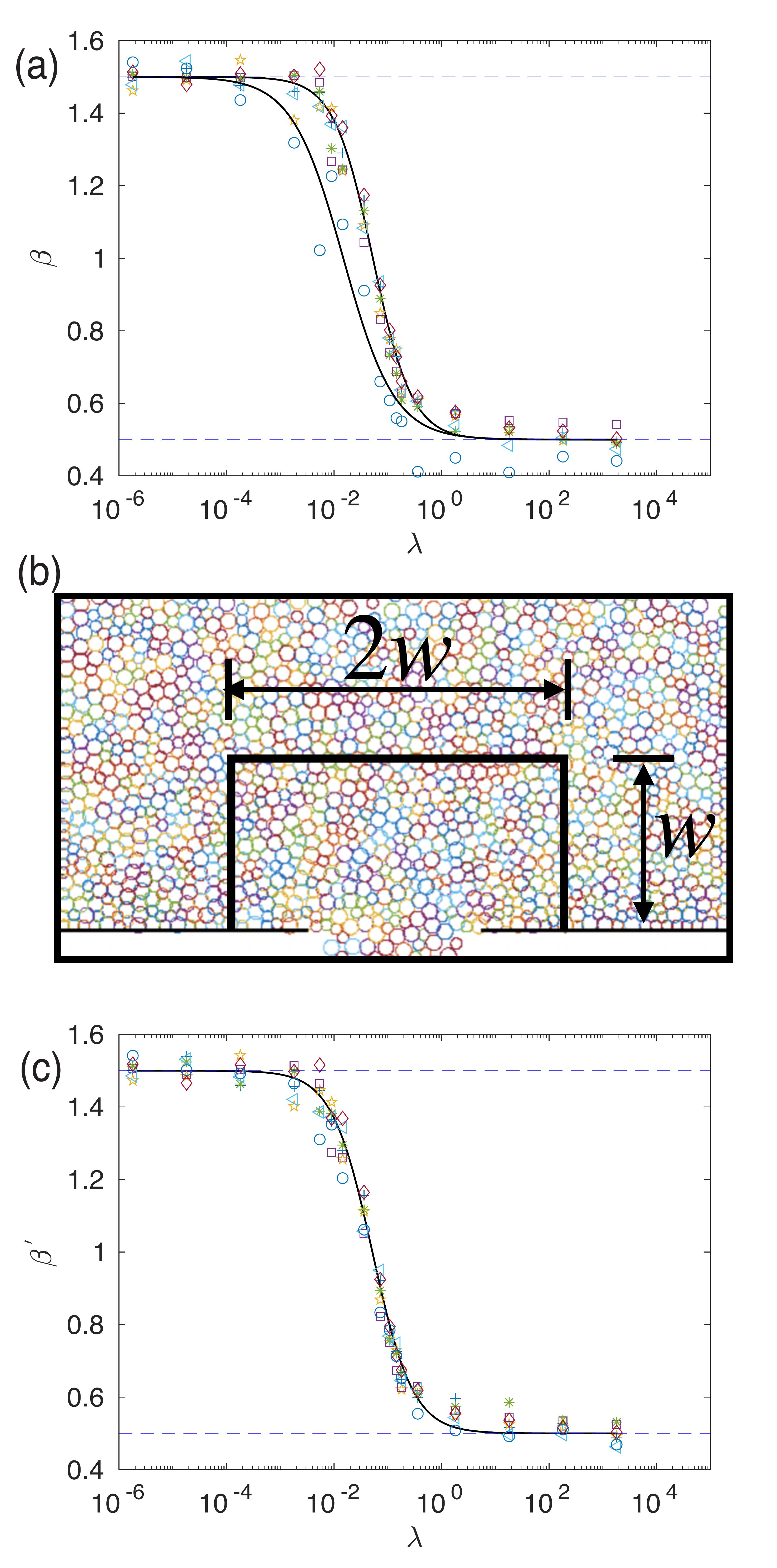}
\caption{(a) Power-law scaling exponent $\beta$ plotted versus the ratio of the viscous drag to the kinetic friction coefficients $\lambda$ from the data in  Fig.~\ref{fig:drag2friction} (a) for the SP and DP models in 2D with $\theta_w =90^{\circ}$. We also show data for the SP model in 2D with $E_{sp}=20$ (circles). (b) Schematic of the $2w \times w$ rectangular region in 2D over which the particle number density $\rho_n$ is measured. (c) The power-law scaling exponent $\beta^{'}(\lambda)$ obtained by fitting the corrected area flow rate, $Q^{'}=Q a_{\rm eff}/a_{\rm avg}$ to Eq.~\ref{eq:1}.} 
\label{fig:spV}
\end{figure}

\section*{Appendix B}
\label{app:B}

In this Appendix, we show results for the area flow rate $Q(w)$ in the rigid-particle limit for both cases $\lambda \rightarrow 0$ and $\lambda \rightarrow \infty$.  We also show that conservation of total particle area is important for accurately modeling the area flow rate in hopper flows of soft and deformable particles in 2D. (Similar results are found in 3D.) As discussed in Sec.~\ref{sim_methods}, the SP model does not explicitly model particle shape change, but instead mimics particle deformability by allowing overlaps between neighboring particles. As a result, the SP model does not conserve total particle area. In contrast, the DP model includes a term in the shape-energy function to conserve particle area as particles change their shapes. (See Eq.~\ref{shape_energy}.) However, in the large-$E_{sp}$ limit for the SP model, where particle overlaps in SP model are small, and in the large-$K_b$ limit for the DP model, the area flow rate $Q(w)$ is same for these two models. As shown in Fig.\ref{fig:rigid}, $Q(w)$ is nearly identical for the SP model with $E_{sp}=10^5$ and for the frictionless DP model with $K_b=10^2$ in the $\lambda \rightarrow 0$ and $\lambda \rightarrow \infty$ limits. For $\lambda \rightarrow 0$ with $\beta= 3/2$, the offset $k \approx 1.6$ in the hard-particle limit, in close agreement with experiments of hopper flows of frictional grains. For $\lambda \rightarrow \infty$ with $\beta= 1/2$, the offset $k \approx 3.4$ in the hard-particle limit. Note that the offset $k$ has different values for $\lambda \rightarrow 0$ and $\lambda \rightarrow \infty$ in the hard-particle limit, which suggests that $k$ is controlled by the flow dynamics and cannot be determined by the hopper geometry alone~\cite{Wilson2014,Fan2022}.

\begin{figure}[!ht]
\includegraphics[width=\columnwidth]{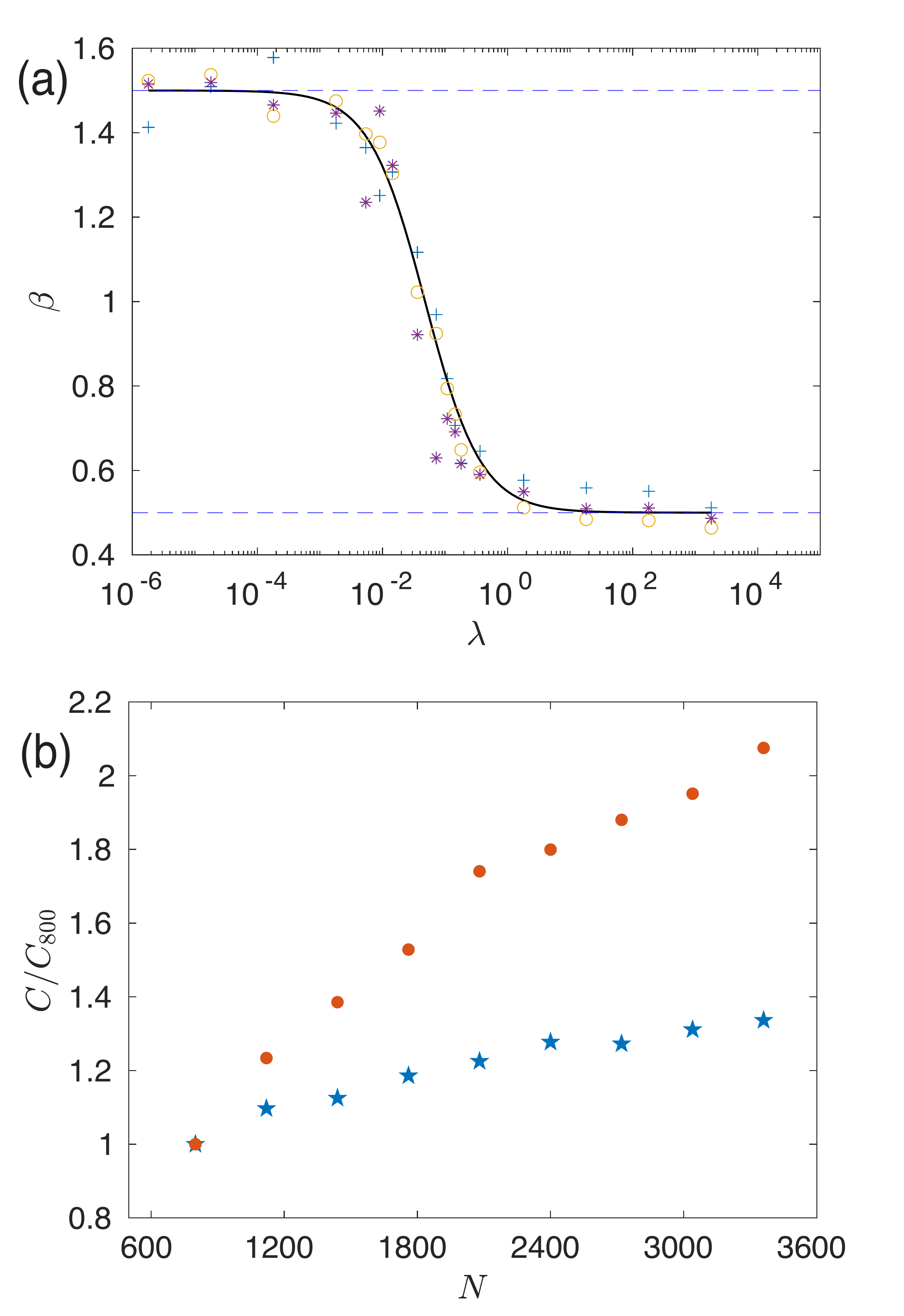}
\caption{(a) Power-law scaling exponent $\beta$ versus the ratio of the viscous drag and the kinetic friction coefficients $\lambda$ for the SP model in 2D with $E_{sp}=50$, $\theta_w = 90^{\circ}$, and $N=800$ (crosses), $1600$ (circles), and $3200$ (asterisks). The horizontal dashed lines indicate $\beta = 3/2$ and $1/2$. The solid line is a fit to Eq.~\ref{eq:b_sig}. (b) Prefactor $C$ in Eq.~\ref{eq:1} normalized by the value at $N=800$ versus the system size $N$ for the 2D SP model for both $\lambda \rightarrow 0$ (circles) and $\lambda \rightarrow \infty$ (stars). } 
\label{fig:sysSize}
\end{figure}

For extremely soft particles, the overlaps that occur in the SP model are sufficiently large that they influence the hopper flow dynamics. For example, in 
Fig.\ref{fig:spV}(a), we show the power-law exponent $\beta$ as a function of $\lambda$ for the SP model with $E_{sp}=20$ in addition to all of the data in Fig.\ref{fig:drag2friction}(a). For this data, the area flow rate $Q$ was calculated by counting the number of mass points that flow past the orifice opening per unit time divided by the particle areal mass density $\rho$. The power-law exponent $\beta(\lambda)$ for the SP model with extremely large overlaps ($E_{sp}=20$) deviates from all of the other data. We can correct $Q(w)$ for the SP model with large particle overlaps by determining the true particle area flowing through the hopper orifice. To do this, we consider a $2w \times w$ rectangular region near the hopper orifice as shown in Fig.\ref{fig:spV}(b) and measure the number density $\rho_n = N/A$ in this region. The effective particle area in this region is $a_{\rm eff}=1/\rho_n$, and the corrected area flow rate is $Q' = Q a_{\rm eff}/a_{\rm avg}$, where $a_{\rm avg} = (a_s + a_l)/2$. In Fig.\ref{fig:spV} (c), we show the power-law scaling exponent $\beta'$ obtained from fitting $Q'$ to Eq.~\ref{eq:1}.  We find that the data from Fig.\ref{fig:drag2friction}(a) (where the particle overlaps are small) do not change and $\beta^{'}=\beta$.  However, $\beta^{'}$ for the SP model with $E_{sp}=20$ shifts so that it falls on the rest of the data from Fig.\ref{fig:drag2friction}(a).

\section*{Appendix C}

In this Appendix, we investigate how the power-law exponent $\beta(\lambda)$ obtained by fitting $Q(w)$ to Eq.~\ref{eq:1} depends on system size for the SP model in 2D. In Fig.~\ref{fig:sysSize} (a), we show $\beta(\lambda)$ for the 2D SP model with $E_{sp}=10^2$ and $\theta_w = 90^{\circ}$ for $N=800$, $1600$, and $N=3200$. We find that $\beta$ is very weakly dependent on system size for the 2D SP model, and we expect similar results for the SP model in 3D. Based on our recent studies of jamming of deformable particles, we expect similar weak system size dependence of $\beta$ for the 2D DP model\cite{Boromand2018,Boromand2019}. We also show the system size dependence of the prefactor $C$ in Eq.~\ref{eq:1} for the 2D SP model in Fig.\ref{fig:sysSize} (b). $C$ grows roughly linearly with system size, but the slope is much weaker for systems with $\lambda \rightarrow \infty$.

\section*{Acknowledgments}
We acknowledge support from NSF Grants No. CBET-2002782 (Y. C., J. D. T., and C. S. O.), No. CBET-2002815 (B. L., P. H., and E. R. W.), and No. CBET-2002797 (M. D. S.). This work was also supported by the High Performance Computing facilities operated by Yale’s Center for Research Computing.

\bibliography{bib}
\bibliographystyle{rsc}
\end{document}